\documentclass{emulateapj} 
 
\newcommand{\lsim}{\raisebox{-0.13cm}{~\shortstack{$<$ \\[-0.07cm] $\sim$}}~}
\newcommand{\gsim}{\raisebox{-0.13cm}{~\shortstack{$>$ \\[-0.07cm] $\sim$}}~}
\newcommand{\egm}{$\rm  8 \, \mu m \,$}
\newcommand{\tfm}{$\rm 24 \, \mu m \,$}
\newcommand{\nLnegm}{$\nu L_\nu^{\rm 8 \, \mu m}$}
\newcommand{\lb}{$L_{\rm bol.}^{\rm IR}\,$}

\shorttitle{The IR luminosity function of galaxies at $\lowercase{z}=1$ and $\lowercase{z} \sim 2$}
\shortauthors{K. I. Caputi et al.}

\begin{document} 
 
 
\title{The infrared luminosity function of galaxies at redshifts $\lowercase{z}=1$ and $\lowercase{z}\sim2$ in the GOODS fields}

 
\author{K. I. \ Caputi\altaffilmark{1,2}, 
G. \ Lagache\altaffilmark{1},
Lin Yan\altaffilmark{3}, 
H. Dole\altaffilmark{1}, 
N. Bavouzet\altaffilmark{1},
E. Le Floc'h\altaffilmark{4}, 
P. I. Choi\altaffilmark{3},
G. Helou\altaffilmark{3} and 
N. Reddy\altaffilmark{5}
} 
\altaffiltext{1} {Institut d'Astrophysique Spatiale, b\^at. 121, 
F-91405 Orsay; Universit\'e Paris-Sud 11 and CNRS (UMR 8617), France}
\altaffiltext{2}{Present address: Institute of Astronomy, Swiss Federal Institute of Technology (ETH H\"onggerberg), CH-8093, Z\"urich, Switzerland. E-mail address: caputi@phys.ethz.ch}
\altaffiltext{3}{Spitzer Science Center, California Institute of Technology, MS 220-6, Pasadena CA 91125, USA}
\altaffiltext{4}{Institute for Astronomy, University of Hawaii, 2680 Woodlawn Drive, Honolulu, HI 96822, USA}
\altaffiltext{5} {Astronomy Option, California Institute of Technology, MS 105-24, Pasadena, 
CA 91125, USA} 



\begin{abstract}
  We present the rest-frame \egm luminosity function (LF) at redshifts $z=1$ and $\sim2$, computed from {\em Spitzer} \tfm-selected galaxies  in the GOODS fields over an area of 291 arcmin$^2$.  Using classification criteria based on X-ray data and IRAC colours, we identify the AGN in our sample. The rest-frame \egm LF for star-forming galaxies at redshifts $z=1$ and $\sim2$ have the same shape as at $z\sim0$, but with a strong positive luminosity evolution.   The number density of star-forming galaxies with $\log_{10}(\nu L_{\nu}^{8 \,\rm \mu m})>11$  increases by a factor $>250$ from redshift $z\sim0$ to 1, and is basically the same at $z=1$ and $\sim2$.   The resulting rest-frame \egm luminosity densities associated with star formation at $z=1$ and $\sim2$ are more than four and two times larger than at $z\sim0$, respectively.    We also compute the total rest-frame \egm LF for star-forming galaxies and AGN at $z\sim2$ and show that AGN dominate its bright end, which is well-described by a power-law. Using a new calibration based on {\em Spitzer} star-forming galaxies at $0<z<0.6$ and validated at higher redshifts through stacking analysis, we compute the bolometric infrared (IR) LF for star-forming galaxies at $z=1$ and $\sim2$. We find that the respective bolometric IR luminosity densities are  $(1.2\pm0.2) \times 10^9$ and $(6.6^{+1.2}_{-1.0}) \times 10^8 \, \rm L_\odot Mpc^{-3}$, in agreement with previous studies within the error bars. At $z\sim2$, around 90\% of the IR luminosity density associated with star formation is produced by luminous and ultraluminous IR galaxies (LIRG and ULIRG), with the two populations contributing in roughly similar amounts. Finally, we discuss the consistency of our findings with other existing observational results on galaxy evolution.  
\end{abstract}



\section{Introduction}
\label{sec-intro}

 Since the {\em Spitzer Space Telescope} (Werner et al.~2004) became operational in December 2003, very important progress has been made in understanding the nature and properties of infrared (IR) galaxies. This progress has been revolutionary, in particular, for the study of galaxies at high redshifts ($z > 1$), to which all of the previous IR facilities operating in the wavelength range $\lambda \sim 5-200 \, \rm \mu m$ had basically no access. Previous missions as the {\em Infrared Astronomical Satellite (IRAS)} and the {\em Infrared Space Observatory (ISO)} allowed to produce multiple studies of mid- and far-IR galaxies, but they were restricted to lower redshifts ($z \lsim 1$) due to their sensitivity limits. Until the launch of {\em Spitzer}, our vision of the high-redshift IR Universe was biased to the relatively small number of galaxies detected in sub-millimetre and millimetre surveys (e.g. Scott et al.~2002; Webb et al.~2003; Greve et al.~2005). 
 
 The sensitivity achieved by the Multiband Imaging Photometer for {\em Spitzer} (MIPS; Rieke et al. 2004) at \tfm is enabling  for the first time to conduct systematic studies of IR galaxies at high redshifts.  Several recent works  have shown that, in contrast to what happens in the local Universe, the IR extragalactic light is increasingly dominated by luminous and ultra-luminous IR galaxies (LIRG and ULIRG, respectively)  with increasing redshift (e.g. Le Floc'h et al.~2004; Lonsdale et al.~2004; Yan et al. 2004; Le Floc'h et al.~2005; Caputi et al.~2006a,b). These LIRG and ULIRG constitute an important fraction of the most massive galaxies present at $z \gsim 1$ (Caputi et al.~2006b).

 In a minor but non-negligible fraction of high-redshift IR galaxies, the IR emission is produced by the presence of an active galactic nucleus (AGN). The exact proportion of AGN-dominated IR galaxies is actually not known, and the determination of such ratio is one of the main problems of IR astronomy. A definitive AGN/star-forming galaxy separation requires the knowledge of the far-IR spectral energy distribution (SED) of these galaxies. Unfortunately, this is not possible for most of high-$z$ galaxies, as their far-IR emission is usually below the confusion limits at far-IR wavelengths (Dole et al.~2004).  This separation is also complicated by the existence of mixed systems, where both star-formation and AGN activity significantly contribute to the IR emission (e.g. Lutz et al.~2005; Yan et al.~2005; Le Floc'h et al.~2006). However, the AGN discrimination is essential to disentangle how much of the IR energy density is associated with star-formation.

 The study of a galaxy luminosity function at different redshifts allows to understand the composition of the extragalactic background as a function of look-back time. The analysis of the changes of the LF with redshift is one of the most direct methods to explore the evolution of a galaxy population. The first studies of the IR-galaxy LF in the local Universe and at low ($z \lsim 1$) redshifts have been based on {\em IRAS} and {\em ISO} data (e.g. Saunders et al. 1990; Xu 2000; Takeuchi, Yoshikawa \& Ishii~2003;  Pozzi et al.~2004; Serjeant et al.~2004; Takeuchi et al. 2006). Using the most recent {\em Spitzer}/MIPS data, Le Floc'h et al.~(2005) analysed in detail the evolution of the IR LF from $z=0$ to $z\sim1$. They found a positive evolution both in luminosity and density between these two redshifts, implying that IR galaxies were more numerous and the IR output was dominated by brighter galaxies at $z \sim 1$ than at $z \sim 0$. The IR galaxy LF at higher redshifts have been explored by other authors (P\'erez-Gonz\'alez et al.~2005; Babbedge et al.~2006).

 Rest-frame \egm luminosities, in particular, are of main relevance for star-forming galaxies as they contain information on  polycyclic aromatic hydrocarbon (PAH) emission.  PAH molecules characterise star-forming regions (D\'esert, Boulanger \& Puget 1990) and the associated emission lines dominate the SED of star-forming galaxies between wavelengths $\lambda=3.3$ and $17 \, \rm \mu m$, with a main bump located around \egm. Rest-frame \egm luminosities have been confirmed to be good indicators of knots of star formation (Roussel et al.~2001; F\"orster-Schreiber et al.~2004; Calzetti et al.~2005) and of the overall star-formation activity of star-forming galaxies (e.g. Wu et al.~2005), except in low-luminosity galaxies with intense ultraviolet (UV) radiation fields (Galliano et al.~2005).

 In this work we compute the rest-frame \egm LF  at redshifts $z=1$ and $z \sim 2$, using \tfm-selected galaxies in the two fields of the Great Observatories Origins Deep Survey (GOODS; Giavalisco et al.~2004). At $z\sim2$, where the fraction of AGN appears to be significant, we analyse separately the LF for star-forming galaxies and for the total IR-galaxy population. The two GOODS fields cover a smaller area than those analysed by some other previous studies of the IR LF. However, they benefit from uniquely deep homogeneous photometric datasets,  ranging from the X-rays to radio wavelengths, as well as an important spectroscopic coverage. As we explain in Section \ref{sec_sample}, this makes possible an almost complete identification of \tfm galaxies down to faint fluxes and the derivation of accurate redshift determinations (cf. also Caputi et al. 2006a,c). These two characteristics are essential for a proper computation of the LF at high redshifts,  without any conclusion relying either on completeness or selection-function corrections.

  The layout of this paper is as follows: in Section \ref{sec_sample}, we describe in detail the selection of our \tfm-galaxy samples at redshifts $0.9<z<1.1$ and $1.7<z<2.3$. In Section \ref{sec_agnsep}, we explain how we perform the separation between star-forming galaxies and AGN within our sample. We compute the rest-frame \egm LF at $z=1$ in Section \ref{sec_lf8z1} and analyse its evolution from $z\sim0$.
In Section \ref{sec_lf8}, we present the rest-frame \egm LF  at  $z\sim2$ and extend the analysis of the evolution up to this high redshift. Later, in  Section \ref{sec_lfbol}, we use a new empirical calibration based on {\em Spitzer} galaxies to obtain the bolometric IR LF at different redshifts.  Finally, in Sections \ref{sec_disc} and \ref{sec_conc}, respectively, we discuss our results and present some concluding remarks. We adopt throughout a cosmology with $\rm H_0=70 \,{\rm km \, s^{-1} Mpc^{-1}}$, $\rm \Omega_M=0.3$ and $\rm \Omega_\Lambda=0.7$.

\section{The IR galaxy sample in the GOODS fields}
\label{sec_sample}

The GOODS fields, namely the GOODS/Chandra Deep Field South (GOODS/CDFS) and GOODS/Hubble Deep Field North (GOODS/HDFN), have been observed by {\em Spitzer} as one of the cycle-1 Legacy Science Programs (P.I. Mark Dickinson). Extended areas of the CDFS and HDFN have also been observed as part of the {\em Spitzer} IRAC and MIPS Guaranteed Time Observers (GTO) programs (P.I. Giovanni Fazio and George Rieke, respectively). 
 
 GOODS/IRAC maps at 3.6 to \egm and MIPS maps at \tfm are now publicly available.  The corresponding GOODS public \tfm catalogues have been constructed using prior positional information from the IRAC 3.6 and $4.5 \, \rm \mu m$ images and by an additional blind extraction of \tfm sources. The resulting \tfm catalogues are  basically reliable and complete for galaxies with fluxes down to  $S(\rm 24 \, \mu m \,)=80 \, \rm \mu Jy$ (Chary et al., in preparation, and cf. the {\em Spitzer} GOODS website\footnote{http://data.spitzer.caltech.edu/popular/goods}). For a comparison, we note that the \tfm catalogue constructed from the shallower MIPS/GTO observations of the CDFS achieves $\sim 80\%$ completeness and only has $\sim 2\%$ of  spurious sources at a similar flux level (Papovich et al.~2004). Although, in principle, fainter sources can be detected in the deeper GOODS images, we decide to only use the conservative GOODS $S(\rm 24 \, \mu m \,)>80 \, \rm \mu Jy$ galaxy catalogues for the selection of our \tfm-galaxy samples at $z\sim1$ and $z\sim2$. In this way, our computed LF are virtually not affected by incompleteness corrections (cf. sections \ref{sec_lf8z1} and \ref{sec_lf8}).

\subsection{Multiwavelength analysis and redshift determinations for \tfm sources in the GOODS/CDFS}

 In the GOODS/CDFS, we restrict our analysis to the 131 arcmin$^2$ which have deep $J$ and $K_s$-band coverage by the Infrared Spectrometer and Array Camera (ISAAC) on the `Antu' Very Large Telescope (Antu-VLT) (GOODS/EIS v1.0 release; Vandame et al., in preparation).  We used the $K_s<21.5$ (Vega mag) galaxy catalogue constructed by Caputi et al.~(2006c) to identify the \tfm galaxies in the GOODS/CDFS catalogue, using a matching radius of 2$^{\prime\prime}$. The percentage of \tfm galaxies with double $K_s$-band identifications within this radius is only $\lsim 8\%$ and 95\% of the associations can be done restricting  the matching radius to 1.5$^{\prime\prime}$ (Caputi et al.~2006b). In all cases of multiple identifications, we considered that the counterpart to the \tfm source was the $K_s$ galaxy closest to the \tfm source centroid. The $K_s<21.5$ mag catalogue allows us to identify 515 \tfm galaxies within the 131 arcmin$^2$ area, i.e. $\sim 94\%$ of the \tfm galaxies with $S(\rm 24 \, \mu m \,)>80 \, \rm \mu Jy$ in this field.

 Caputi et al.~(2006c) measured multiwavelength photometry for all their $K_s<21.5$ mag galaxies. They ran SEXTRACTOR  (Bertin \& Arnouts 1996) in `double-image mode' to perform aperture photometry on the GOODS/EIS v1.0 $J$-band images, centred at the position of the $K_s$-band extracted sources. They also looked for counterparts of the $K_s<21.5$ mag sources in the public GOODS Advanced Camera for Surveys (ACS) catalogues, which provided photometry in the $B, V, I_{775}$ and $z$ bands. The stellarity parameter measured on the $z$-band images allowed them to separate out galactic stars. Finally, they ran SEXTRACTOR on the Spitzer/IRAC  $\rm 3.6$  and  $\rm 4.5 \, \mu m$  images to identify the $K_s<21.5$ mag galaxies and measured aperture photometry at these longer wavelengths. We refer the reader to Caputi et al.~(2006c) for additional details about the photometric measurements and applied aperture corrections.

  Caputi et al.~(2006c) obtained an estimated redshift for each one of their galaxies modelling their stellar SED  from the B through the $\rm 4.5 \, \mu m$ bands. They used the public code HYPERZ (Bolzonella, Miralles \& Pell\'{o}, 2000) with the GISSEL98  template library (Bruzual \& Charlot 1993) and the Calzetti et al.~(2000) reddening law to account for internal dust extinction. 
  
   The HYPERZ redshift estimates have been replaced by COMBO17 photometric redshifts (Wolf et al.~2004) for those galaxies with magnitudes  $R<23.5$ mag at redshift $z<1$, which is the regime of higher accuracy for COMBO17. In these cases, the SED fitting has been constrained to the COMBO17 redshifts. The cross-correlation of the GOODS/CDFS  \tfm catalogue with the Caputi et al.~(2006c) $K_s<21.5$ mag catalogue  directly gives us estimated redshifts and best-fitting SED models for all the identified \tfm galaxies.

\subsection{Multiwavelength analysis and redshift determinations for \tfm sources in the GOODS/HDFN}

 In the GOODS/HDFN, we followed a similar strategy for the analysis of sources as in the GOODS/CDFS. However, unfortunately, we only have access to deep $K_s$-band data for a part of this field (Reddy et al.~2006b). Thus, we used the Spitzer/IRAC  $\rm 3.6 \, \mu m$ maps to identify the \tfm galaxies. We analysed in this case the entire GOODS/HDFN region, i.e. the 160 arcmin$^2$ with deep GOODS/HST-ACS coverage. We ran SEXTRACTOR on the IRAC $\rm 3.6$  and  $\rm 4.5 \, \mu m$  images. We constructed a catalogue of $\rm 3.6 \, \mu m$ sources, accepting only those objects also identified in the $\rm 4.5 \, \mu m$ band. To encompass the technique applied by Caputi et al. (2006c) on the IRAC maps of  the GOODS/CDFS, we measured  photometry in  circular apertures of 2.83$^{\prime\prime}$-diameter\footnote{The aperture size has been chosen in correspondence to the aperture sizes used in the GOODS ACS catalogues.} and applied  aperture corrections of 0.50 and 0.55 mag to the $\rm 3.6$  and  $\rm 4.5 \, \mu m$ magnitudes, respectively. We then used this $\rm 3.6 \, \mu m$ catalogue to identify the \tfm sources in the GOODS/HDFN, using a matching radius of 2$^{\prime\prime}$. This allows us to  identify 856 \tfm galaxies in the 160 arcmin$^2$ of the GOODS/HDFN, i.e. $\sim 95\%$ of the \tfm galaxies with $S(\rm 24 \, \mu m)>80 \, \rm \mu Jy$ in this field. The identification completeness achieved for \tfm galaxies in this field using $\rm 3.6 \, \mu m$ sources is similar to the identification completeness obtained for \tfm galaxies in the GOODS/CDFS using $K_s$-band sources. This indicates that the two identification methods are basically equivalent. In any case, the IRAC 3.6 and  $\rm 4.5 \, \mu m$ data are incorporated in the  SED modelling of all the sources in the two fields.

 We followed up in the optical bands those IRAC $\rm 3.6 \, \mu m$ objects which were counterparts to $S(\rm 24 \, \mu m)>80 \, \rm \mu Jy$ sources. Once more, we used the public GOODS ACS catalogues  to obtain aperture photometry in the $B, V, I_{775}$ and $z$ bands.   In addition, we looked for counterparts of the  $\rm 3.6 \, \mu m$ sources in the U and HK'-band images of the GOODS/HDFN (Capak et al.~2004). Although these images are relatively shallower than the other optical/near-IR data available for this field, we decided to include these data to improve the SED coverage. Finally, we incorporated the deep $J$ and $K_s$-band data from Reddy et al.~(2006b) for those galaxies lying in the region where these data were available ($< 40\%$ of the analysed area).

 We used the multiwavelength data from the U to the  $\rm 4.5 \, \mu m$ bands to model the SED and obtain photometric redshifts for all of our \tfm galaxies in the GOODS/HDFN using  HYPERZ, in an analogous way to that in Caputi et al.~(2006c). As in the latter, we applied a set of criteria to control the HYPERZ output: 1) the photometric redshifts for galaxies detected in the shallow U-band catalogues were constrained to  a maximum value $z_{phot}=2$, as bright U-band sources are unlikely to be beyond these redshifts; 2) analagously, the estimated redshifts of galaxies not-detected in the U-band but detected in the B-band were constrained to a maximum value $z_{phot}=4$;  3) for the GOODS/HDFN catalogue, we found that HYPERZ produced an overdensity of galaxies in the redshift range $1.5<z_{phot}<1.7$. Comparison with spectroscopic redshifts (see below) suggested that this overdensity was an artefact of HYPERZ applied to our sample. Thus, to test these possible spurious redshifts, we double-checked the fitting of all the galaxies with HYPERZ redshift $1.5<z_{phot}<1.7$  using the P\'EGASE library (Le Borgne \& Rocca-Volmerange 2002). We kept the HYPERZ solution for those sources confirmed by P\'EGASE as belonging to the 1.5-1.7 redshift range. For all the remaining $1.5<z_{phot}<1.7$ galaxies, we replaced the photometric redshift by the P\'EGASE estimate.   This strategy improved the agreement with spectroscopic redshifts. The percentage of galaxies with P\'EGASE redshifts in our final \tfm catalogue for the GOODS/HDFN is 5\%.

\subsection{The final IR galaxy samples in the combined GOODS fields} 
\label{sec_finsamp}

 Our final \tfm catalogue contains 1371  \tfm sources with $S(\rm 24 \, \mu m)>80 \, \rm \mu Jy$ over a total area of 291 arcmin$^2$. We identified only 22 out of 1371 sources as galactic stars. All the remaining sources are galaxies. Our aim is to separate two sub-samples of galaxies from this final catalogue: 1) the \tfm  galaxies with redshifts $0.9<z<1.1$ for the computation of the IR LF at $z=1$; 2) the \tfm  galaxies with redshifts $1.7<z<2.3$ for the computation of the IR LF at $z\sim2$.

\begin{figure}
\plotone{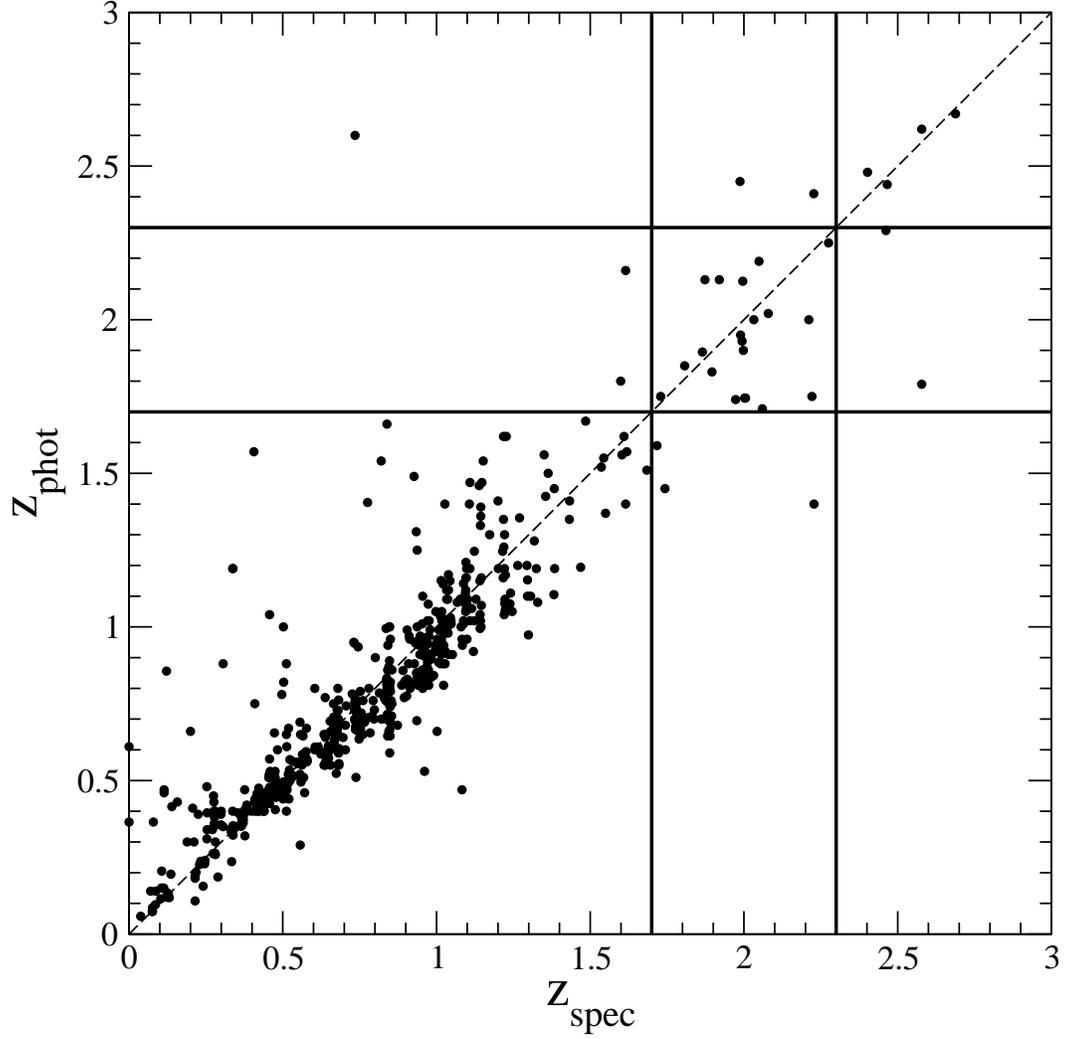}
\caption[]{\label{fig_zphzsp} The comparison between photometric and spectroscopic redshifts for galaxies in our \tfm-selected sample in the GOODS fields. The distribution of relative errors $dz=(z_{phot}-z_{spec})/(1+z_{spec})$ has a median -0.007 and a dispersion $\sigma_z=0.05$. The horizontal lines separate the galaxies with $1.7<z_{phot}<2.3$ and the vertical lines, those with $1.7<z_{spec}<2.3$.  The distribution of relative errors for the $1.7<z_{spec}<2.3$ subsample of galaxies has a median -0.01 and a dispersion $\sigma_z=0.06$.}
\end{figure}

 We performed a final step before separating the two definitive sub-samples of \tfm  galaxies used in this work. In addition to the wealth of photometric data, both GOODS fields benefit from an important amount of spectroscopic data, most of which are publicly available (Cohen et al.~1996; Le F\`evre et al.~2004; Wirth et al.~2004;  Vanzella et al.~2005, 2006; Choi et al., in preparation; among others). Some additional redshifts in the GOODS/CDFS have been kindly made available to us by Fran{\c c}ois Hammer and H\'ector Flores. We compiled these data and found that more than $45\%$ of our \tfm galaxies in  the combined fields had spectroscopic redshifts. We incorporated these spectroscopic redshifts into our catalogue, which superseded the corresponding photometric values. The finally discarded photometric redshifts have been used to assess the quality of our redshift estimates. Figure \ref{fig_zphzsp} shows the comparison between photometric and spectroscopic redshifts for the galaxies in our sample for which both redshifts are available. We observe a good agreement between photometric estimates and real redshifts. The distribution of relative errors $dz=(z_{phot}-z_{spec})/(1+z_{spec})$ has a median value -0.007 and the dispersion is $\sigma_z=0.05$.

 From the definitive redshift catalogue which incorporates spectroscopic redshifts, we select those \tfm galaxies lying at $0.9<z<1.1$ and  $1.7<z<2.3$. 
 
 The $0.9<z<1.1$ sample is composed of 227 galaxies with  $S(\rm 24 \, \mu m)>80 \, \rm \mu Jy$ and a median redshift $z=1.00$. We use this sample to compute the IR LF at $z=1$. More than 60\% of these galaxies have spectroscopic redshifts $z_{\rm spec}$. The quality of photometric redshifts is similar as that for the total sample: the median of relative errors $dz=(z_{phot}-z_{spec})/(1+z_{spec})$  is -0.01 and the dispersion is $\sigma_z=0.05$. In the computation of the  IR LF at $z=1$, we consider that these errors only affect those galaxies with photometric redshifts ($<40\%$).

 Our $1.7<z<2.3$ sample contains 161 \tfm galaxies with  $S(\rm 24 \, \mu m)>80 \, \rm \mu Jy$. This is the sample we use to compute the IR LF at redshift $z\sim2$. The median redshifts of these 161 galaxies is $z=1.93$. Although for practicity we refer to these galaxies as the $z\sim2$ sample, all the calculations made in Sections \ref{sec_lf8} and \ref{sec_lfbol} take into account the actual median redshift value. More than 15\% of the galaxies selected with $1.7<z<2.3$ have spectroscopic redshifts. 
The quality of photometric redshifts for the $z\sim2$ sample can also be assessed from figure  \ref{fig_zphzsp}.   We see that the agreement between photometric and spectroscopic redshifts is still very reasonable for this high-redshift sample.  The distribution of relative errors $dz=(z_{phot}-z_{spec})/(1+z_{spec})$ has a median -0.01 and a dispersion $\sigma_z=0.06$. This statistics has been computed based on all sources (i.e. AGN included; see below). This suggests that the SED templates we use to derive photometric redshifts are suitable for all our sample. The photometric redshift error bars affect the majority of galaxies in our $z\sim2$ sample and are taken into account in the computation of the corresponding LF, as we explain in Section \ref{sec_lf8}.
 
\begin{figure}
\plotone{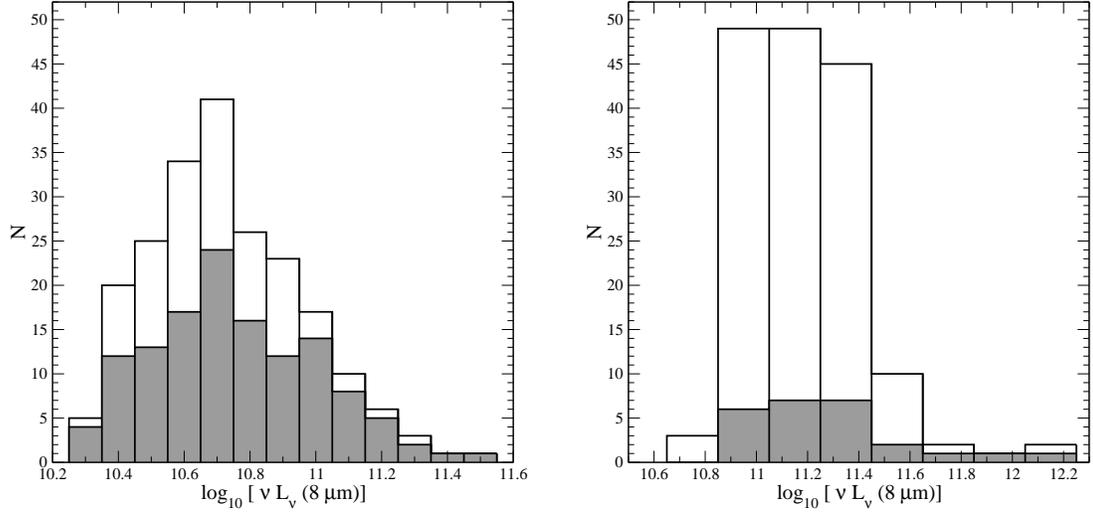}
\caption[]{\label{fig_l8histo} The distribution of rest-frame \egm luminosities for galaxies at redshifts $0.9<z<1.1$ (left-hand panel) and $1.7<z<2.3$ (right-hand panel). In each panel, the empty and shaded  histograms include all the galaxies and only those with spectroscopic redshifts, respectively.}
\end{figure}

  We note that the galaxies with spectroscopic redshifts are representative of our  entire \tfm sample in each of the considered redshift bins ($0.9<z<1.1$ and $1.7<z<2.3$). The two panels in Figure \ref{fig_l8histo} show the rest-frame \egm luminosities of all of our galaxies (empty histograms) and those of galaxies with spectroscopic redshifts (shaded histograms), at these different redshifts. Details on the calculation of \egm luminosities are given in Section \ref{sec_lf8z1}. From Figure \ref{fig_l8histo}, we can see that galaxies with spectroscopic redshifts basically span the whole range of IR luminosities considered in this work. Thus, the errors derived from the comparison of photometric and spectroscopic redshifts are applicable to the entire IR LF. 
  
\section{The normal/active galaxy separation}
\label{sec_agnsep}

 In this work, we would like to compare the IR LF for star-forming galaxies only with the total IR LF. To do this, we need to identify the active galaxies present in our sample.

\begin{figure} 
\plotone{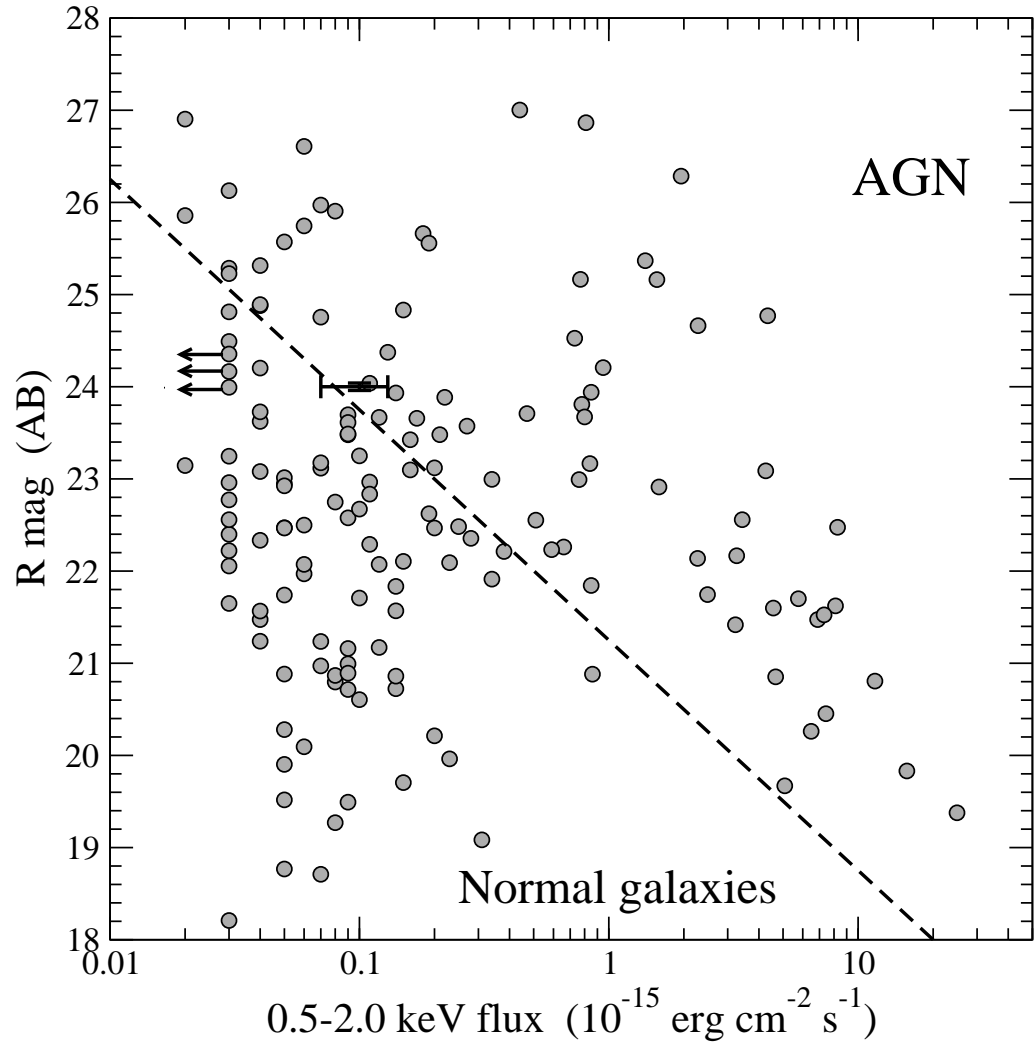}
 \caption[]{\label{fig_rvsx} $R$-band magnitudes versus soft X-ray fluxes for the X-ray detected galaxies in our \tfm galaxy sample in the GOODS/HDFN. The error bar for a generic source with  soft X-ray flux $10^{-16}\, \rm erg \, cm^{-2}s^{-1}$ and $R=24$ mag is shown. The left-pointing arrows indicate that the soft X-ray flux  $3\times10^{-17}\, \rm erg \, cm^{-2}s^{-1}$ is an upper limit.}
\end{figure}

 One of the most efficient ways of identifying AGN is through their X-ray emission. The GOODS fields have deep X-ray coverage obtained with the {\em Chandra X-ray Observatory}: the 1Ms maps for the CDFS (Giacconi et al. 2002) and the 2Ms maps for the HDFN (Alexander et al. 2003).  We used the corresponding public X-ray catalogues to identify the AGN within our sample. However, given the depth of these catalogues (especially that of the HDFN), X-ray sources include not only quasars and AGN but also powerful starburts which also emit in X-rays. To separate the two classes of X-ray sources, an optical versus X-ray flux diagram can be used. Figure \ref{fig_rvsx} shows the $R$-band magnitude versus the soft X-ray flux of the X-ray-detected galaxies in our \tfm sample in the HDFN. The $R$-band magnitudes of our galaxies have been interpolated using the $V$ and $I_{775}$ magnitudes. This plot is similar to that presented in Alexander et al.~(2003). The dashed line shows the empirical separation between normal galaxies and AGN, as calibrated by Hornschemeier et al.~(2001). Using this diagram, we identify the X-ray detected AGN within our \tfm sample and, in particular,  those at $0.9<z<1.1$ and $1.7<z<2.3$.

  Some AGN with weak soft X-ray fluxes but significant emission in the hard bands can contaminate the normal galaxy region in the 
$R$-band magnitude versus soft X-ray flux diagram. These AGN are characterised by a flat photon index $\Gamma<1.0$ (e.g. Hornschemeier et al.~2003). We also looked for these kinds of objects to identify the AGN present in our sample.

\begin{figure}
\plotone{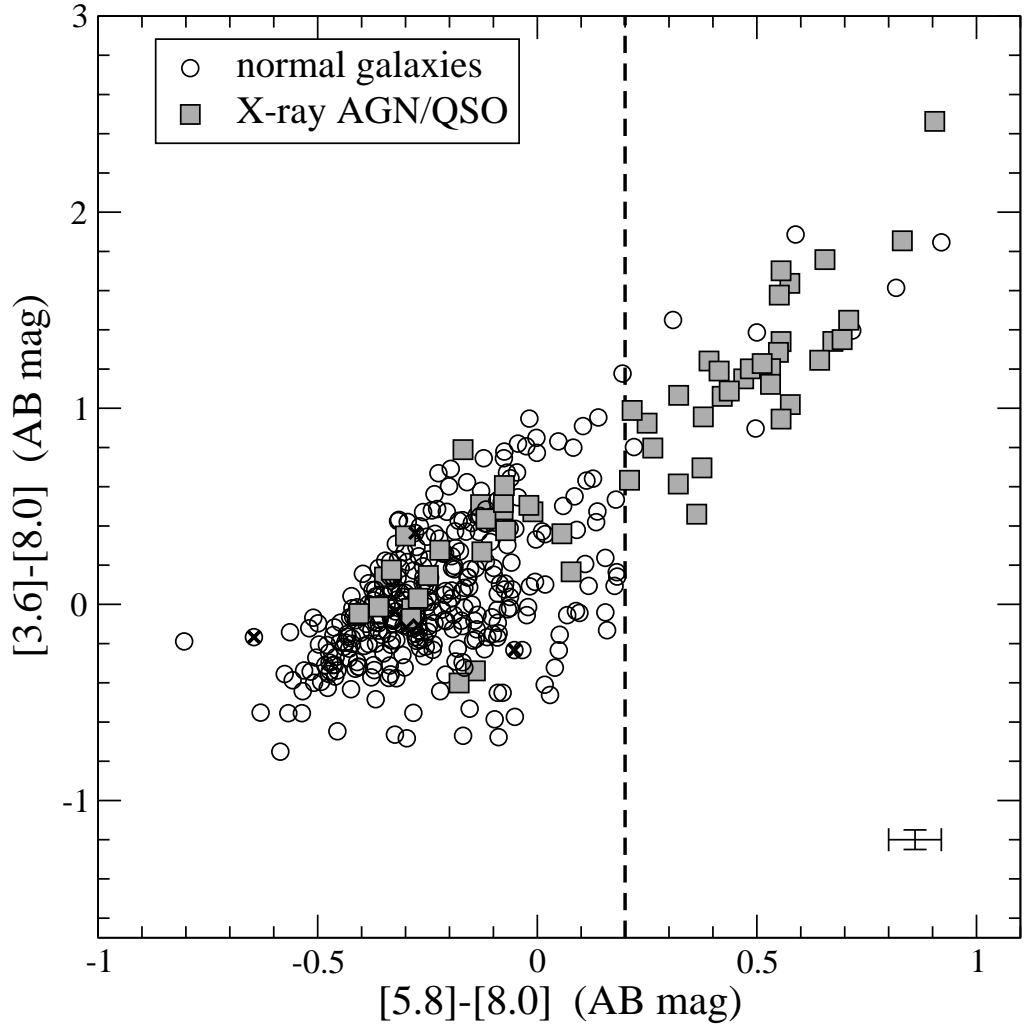}
\caption[]{\label{fig_irac} IRAC-based colour-colour diagram for the \tfm sources with redshifts $z>1.5$ in the GOODS fields. Filled squares and empty circles refer to X-ray classified AGN and to all the other $z>1.5$ \tfm galaxies, respectively. The cross-like symbols indicate the few star-forming galaxies at $z>1.5$ which are X-ray detected. The typical error bars for the colours of these sources are indicated in the lower right corner of the plot.}
\end{figure}

 It is known, however, that the X-ray selection can be incomplete for the selection of AGN. Other active galaxies exist, which are not detected even in deep X-rays surveys. A complementary method to select active galaxies can be developed based on the analysis of the IR colour excess in the {\em Spitzer} IRAC bands. Figure \ref{fig_irac} shows  the (3.6-$\rm  8 \, \mu m$) versus (5.8-$\rm  8 \, \mu m$) colours for all the galaxies with redshift $z>1.5$ in our \tfm sample. Empty circles correspond to all those galaxies not classified as AGN using X-ray data (either not detected in X-rays or X-ray sources classified as starbursts). Filled squares indicate the X-ray classified AGN.
 We restrict this diagram to high redshift sources for the following reason. The stellar bump centred at rest-frame wavelength $\lambda\sim1.6 \,\rm \mu m$ is shifted into the IRAC bands at $z\gsim1.5$. For active galaxies, the galaxy SED at the same rest-frame wavelengths is dominated by a power-law continuum. Thus, it is expected that an IRAC-based colour-colour diagram is able to separate the AGN through their IR excess.   At low redshifts, this separation is much less clear, especially because star-forming galaxies with PAH  emission can mimic the IR excess.  Similar colour-colour plots have been used with the purpose of separating  normal and active galaxies elsewhere (e.g. Lacy et al. 2004; Stern et al. 2005; Caputi et al. 2006b).

 Inspection of figure \ref{fig_irac} shows that X-ray-selected AGN display a wide range of (3.6-$\rm  8 \, \mu m$) and (5.8-$\rm  8 \, \mu m$) colours, while the vast majority of "normal galaxies" (i.e. non X-ray-classified AGN) appear on the left-hand side of this diagram, with a colour  (5.8-$\rm  8 \, \mu m$) $\lsim 0.2$ (AB). As we mentioned above, the relatively blue colours are produced by the stellar SED bump mapped at the IRAC wavelengths. The galaxies lying on the right-hand side, on the contrary, present an excess in the SED continuum which is characteristic of AGN. Thus, based on this diagram, we adopt an empirical colour cut to produce an additional AGN selection criterion: all the $z>1.5$ galaxies with  (5.8-$\rm  8 \, \mu m$) $> 0.2$ (AB) within our sample are classified as AGN. This same additional AGN selection criterion has been used by Caputi et al.~(2006b).

 We would like to note that, while this colour cut produces a safe  criterion to select additional active galaxies, it is possibly not complete. The dispersion of colours displayed by X-ray-selected AGN suggests that other active sources --not detected in X-rays and with no IRAC colour excess-- could also exist among the \tfm galaxies. On the other hand, some of the X-ray classified AGN could be composite systems, where a fraction of the bolometric IR luminosity is actually due to star formation. Unfortunately, no AGN selection criterion appears to be both complete and reliable at the same time (cf. e.g. Barmby et al.~2006). As we do not have information on the far-IR emission of our galaxies, our separation criteria are possibly the most adequate to discriminate AGN.

 For our sample of 227 \tfm galaxies with redshift $0.9<z<1.1$, only the X-ray criteria have been applied. We identify  23 out of 227 galaxies as AGN, i.e. $\sim 10\%$ of the sample. We will exclude the AGN from our sample in order to determine the IR LF for star-forming galaxies at $z=1$, but we note that the inclusion of AGN only has a minor impact on the LF at this redshift. 
 
 For the sample at redshifts $1.7<z<2.3$, we applied both selection criteria to separate AGN (X-ray and IRAC-colour classifications). The fraction of active galaxies at these redshifts appears to be more important than at $z\sim1$. We identify 29 AGN among our 161 \tfm galaxies at $1.7<z<2.3$, i.e. $\sim 18\%$ of the sample. 23 out of these 29 AGN have been identified using X-rays and the remaining 6 AGN have been classified through their IRAC colours.  As we will see below, the LF computed including and excluding AGN have non-negligible differences, because these objects dominate the bright end of the IR LF at these high redshifts. Throughout this paper, when we refer to the star-forming galaxies at redshift $z\sim2$, we mean our sample of 161-29=132 objects which we have not classified as AGN at these redshifts.

\section{The rest-frame \egm LF at redshift $\lowercase{z}=1$}
\label{sec_lf8z1}

\subsection{The k-corrections from $11.4-12.7$ to \egm}
\label{sec_kcz1}

 Before computing the rest-frame \egm LF at $z\sim2$, we aim to understand its evolution from $z\sim0$ to redshift $z=1$. For this, we compute the rest-frame \egm LF for our 204 \tfm-selected star-forming galaxies in the redshift range $0.9<z<1.1$. AGN have been excluded from this analysis.  AGN constitute   $\sim 10\%$ of our sample with $0.9<z<1.1$ and their exclusion does not significantly change the shape of the \egm LF at $z=1$. This is in contrast to what we find at $z\sim2$, where AGN constitute a somewhat higher fraction of sources which dominate the bright end of the rest-frame \egm LF (cf. Section \ref{sec_lf8}).

 We compute the rest-frame \egm luminosity ($\nu L_{\nu}^{8 \,\rm \mu m}$)  of each galaxy as \\$\nu L_{\nu}^{8 \,\rm \mu m}=$ $\nu \, 4 \pi k(\lambda_{\rm rf}) \, S(\rm 24 \, \mu m) \, \mathnormal d^2_{\rm L}(\mathnormal z)$, where $S(\rm 24 \, \mu m)$ is the \tfm flux, $d_{\rm L}(\mathnormal z)$ is the luminosity distance and $k(\lambda_{\rm rf})$, the corresponding k-correction at the rest-frame wavelength $\lambda_{\rm rf}$. The width of the redshift bin we consider $0.9<z<1.1$ implies that the observed \tfm maps rest-frame wavelengths $11.4<\lambda_{\rm rf} < 12.7 \, \rm  \mu m$. We need then to apply k-corrections to convert the rest-frame $11.4-12.7 \, \rm  \mu m$ into \egm fluxes.

\begin{figure}
\plotone{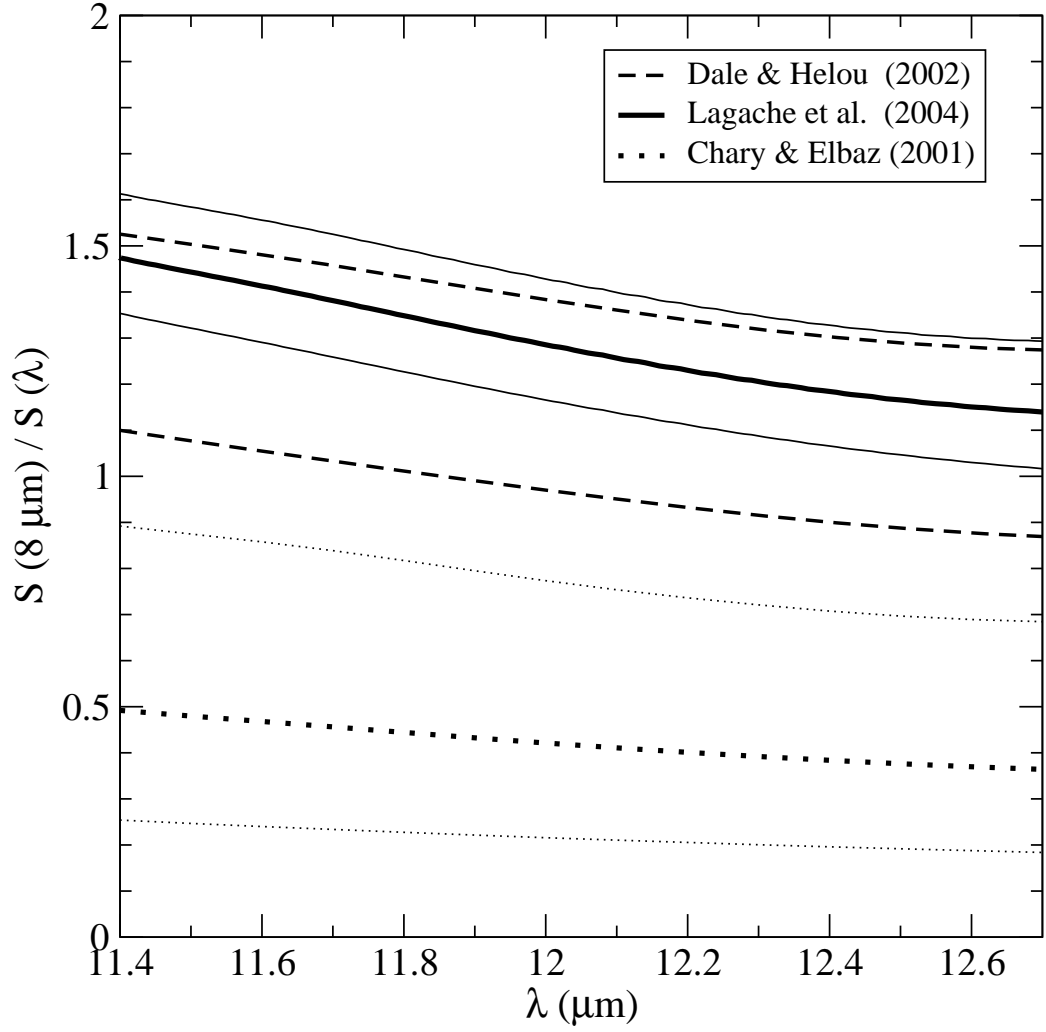}
\caption[]{\label{fig_12to8k} The k-corrections between 11.4-12.7 and \egm fluxes obtained using different IR galaxy model templates: Lagache et al.~(2004; solid lines); Chary \& Elbaz (2001; dotted lines) and Dale \& Helou (2002; dashed line). The thin solid (dotted) lines indicate the interval of corrections obtained using the different models of Lagache et al. (Chary \& Elbaz) with bolometric IR luminosity $L_{\rm IR}>10^{11} \, \rm L_\odot$. The corresponding thick lines indicate median k-corrections. The dashed line corresponds to the Dale \& Helou model with parameter $\alpha=1.1$ and 1.4.}
\end{figure}

 To compute these k-corrections, we analyse different sets of IR galaxy templates available in  the literature, namely the models by Chary \& Elbaz (2001) and Elbaz et al.~(2002); Dale et al.~(2001) and Dale \& Helou~(2002); and Lagache et al.~(2004). We convolve the SED templates in all these models with the transmission function of the \tfm filter and obtain the relation between the fluxes at $11.4-12.7$ and $8 \, \rm \mu m$. Figure \ref{fig_12to8k} shows the $\lambda$-to-\egm k-corrections in the wavelength range $\lambda_{\rm rf}=11.4-12.7 \, \mu \rm m$. Different line styles indicate the k-corrections obtained with different SED templates. The solid and dotted lines correspond to the range of k-corrections derived for galaxies with bolometric IR luminosities $L_{\rm IR}> 10^{11} \, \rm L_\odot$, from the  Lagache et al. and Chary \& Elbaz models, respectively (with thick lines indicating the median values). The dashed lines show the k-corrections obtained with the Dale et al. SED model with parameters $\alpha=1.1$ and 1.4 (cf. Dale et al.~2001). It is clear from inspection of figure \ref{fig_12to8k} that the  k-corrections between $11.4-12.7$ and \egm obtained with these different models have some significant dispersion. These differences are produced by the limited knowledge on PAH emission when modelling the PAH-dominated region of a star-forming galaxy SED.

 In this work, we adopt the median k-corrections obtained with the Lagache et al.~(2004) models of star-forming galaxies with bolometric IR luminosities $L_{\rm IR}> 10^{11} \, \rm L_\odot$ (thick solid line in \ref{fig_12to8k}). As we show in Section \ref{sec_compl8lb}, the Lagache et al. templates produce an \egm-to-bolometric IR luminosity conversion quite close to that measured on the observed SED of {\em Spitzer} galaxies (Bavouzet et al.~2006). This suggests that these templates incorporate an adequate modelling of the PAH emission region in the star-forming galaxy SED.

\subsection{The $1/V_{\rm max}$ method} 
\label{sec_vmaxz1}

   We compute the  rest-frame \egm LF for star-forming galaxies at redshift $z=1$ using the $1/V_{\rm max}$ method (Schmidt 1968). For this calculation, we consider the 204 star-forming galaxies with redshift  $0.9<z<1.1$ within our sample. The advantage of the $1/V_{\rm max}$ technique is that it allows to compute the LF directly from the data, with no parameter dependence or model assumption. Besides, the normalization of the LF is directly obtained from the same calculation. The comoving volume $V_{\rm max}=V_{z_{\rm max}}-V(z=0.9)$ for each source corresponds to the maximum redshift $z_{\rm max}$ at which it would be included in the catalogue, given the limiting flux $S(\rm 24 \, \mu m)=80 \, \rm \mu Jy$, and provided that this redshift is smaller than the maximum of the considered redshift bin (in this case $z=1.1$). Otherwise, $V_{\rm max}$ is equal to the volume corresponding to the $0.9<z<1.1$  bin $V_{\rm max}=V_{\rm bin}$.

   As we explained in Section \ref{sec_sample}, the GOODS \tfm catalogues are basically complete down to the limiting flux and, thus, no sample completeness corrections are needed for our catalogues. However, we do apply completeness corrections to account for the percentage (5-6\%) of unidentified \tfm sources (cf. Section \ref{sec_sample}). These identification completeness corrections are very small and none of the  conclusions presented here depend on the application of such corrections.

   We present the results of our rest-frame \egm LF at redshift $z=1$ computed with the $1/V_{\rm max}$ method in Figure \ref{fig_lf8z1} (upward-pointing triangles). This LF, as well as all the other presented in this work, have been computed jointly on the GOODS/CDFS and GOODS/HDFN. Although we have checked the consistency within the error bars of the LF obtained in the two fields separately,  the sample variance effects are more important than when considering both fields combined (cf. Figure \ref{fig_lf8separa}).    We show the \egm LF function computed with the $1/V_{\rm max}$  method only in the completeness region of \egm luminosities ($\nu L_{\nu}^{8 \,\rm \mu m} \gsim 3 \times 10^{10}\, \rm L_\odot$), imposed by the flux limits of the \tfm survey and the considered redshifts.  The total comoving volume probed by our survey is $1.3\times 10^5 \, \rm  Mpc^3$.

   The error bars for this LF values depend not only on the number of sources (Poisson statistics), but also on the errors in the photometric redshifts and in the k-corrections applied. The errors in the photometric redshifts affect only $<40\%$ of our galaxies  at $0.9<z<1.1$, given the high percentage of available spectroscopic redshifts.   To account for the errors in the photometric redshifts, we made Monte Carlo simulations of our $\nu L_{\nu}^{8 \,\rm \mu m}$ catalogues. We produced 1000 simulated catalogues, each one with the same number of sources as our original $0.9<z<1.1$  catalogue of star-forming galaxies (i.e. 204 sources each). The redshift of each source has been allowed to randomly vary following a Gaussian distribution centred at $z_{\rm centre}=\mathnormal z-0.01$ and with a dispersion $\sigma_z=0.05 \, (1+z)$ (cf.  Section \ref{sec_finsamp}), where $z$ is the redshift of the source in the original catalogue. The redshift of those sources with spectroscopic determinations have been left unchanged. For the k-corrections, we fixed the error bars to $\epsilon=0.50$, which is roughly the dispersion between the different Lagache et al. and Dale \& Helou model predictions (see Figure \ref{fig_12to8k}). To include these errors in the simulations, we computed the rest-frame \egm luminosity $\nu L_{\nu}^{8 \,\rm \mu m}$ of each galaxy in the mock catalogue allowing the corresponding k-correction to have a random value within the range of its error bar. Finally, the LF has been recomputed with the $1/V_{\rm max}$ method for each of the mock catalogues, with exactly the same procedure as for the original catalogue. From the distribution of the LF values in each  $\nu L_{\nu}^{8 \,\rm \mu m}$ bin, we determined the error bars on our original $1/V_{\rm max}$ results.
  
  For a comparison,  we also show the \egm LF of star-forming galaxies at redshift $z\sim0$ (strictly $0<z<0.3$; with median $z\approx0.2$), computed  by Huang et al.~(2006), using the $1/V_{\rm max}$ method applied to IRAC \egm GTO data (cross-like symbols in Figure \ref{fig_lf8z1}). No error bars have been plotted for this LF, as they are significantly smaller than the error bars of the LF we determine here. The comparison of this $1/V_{\rm max}$ LF with our own determination at $z=1$ shows a substantial increment of the density of  star-forming galaxies with rest-frame \egm luminosities $\log_{10}(\nu L_{\nu}^{8 \,\rm \mu m}) \gsim 10.5$, with increasing redshift.  We note that this behaviour is evident from the $1/V_{\rm max}$ calculation, independently of the parametric analysis we discuss below.

\subsection{The maximum likelihood analysis} 
\label{sec_mlz1}

\begin{figure}
\epsscale{0.60}
\plotone{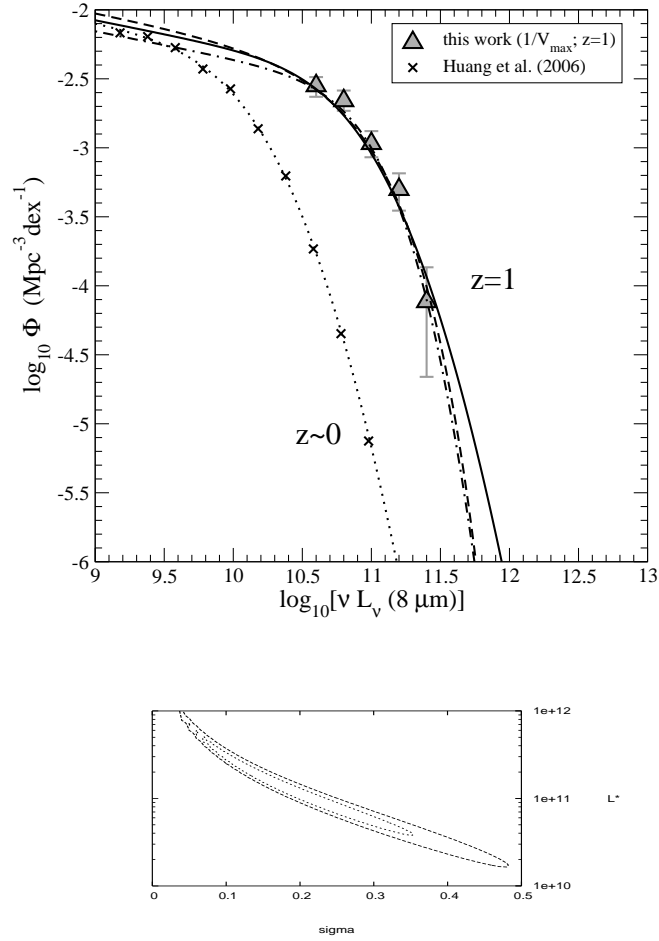}
\caption[]{\label{fig_lf8z1} Upper panel: The rest-frame \egm LF for star-forming galaxies at $z=1$ in the GOODS fields, compared to the \egm LF at $z\sim0$. The cross-like symbols  show the \egm LF for star-forming galaxies at $z\sim0$, as computed by Huang et al.~(2006) with the $1/V_{\rm max}$ method. The dotted line represents the best  $\chi^2$-fit obtained using a double-exponential function as  that in eq. (\ref{eq-dexp}). The upward-pointing triangles show the $1/V_{\rm max}$ LF at $z=1$ obtained in this work, only strictly in the region of completeness of \egm luminosities.   Lines of different styles show  the \egm LF at $z=1$ computed with the ML STY analysis, assuming different laws:  a  double-exponential form  with bright-end slope fixed to the local value ($\sigma=0.36$; solid line); the same double-exponential form with a free $\sigma$ parameter (dotted-dashed line), and a Schechter function (dashed line).
Lower panel: the 68.3 and 95.4\% confidence levels in $(\sigma, \nu L_{\nu}^{\ast \, 8 \,\rm \mu m})$ space in the case  of a double-exponential law with $\sigma$ as a free parameter. The parameter values yielding the ML are $\sigma=0.20^{+0.11}_{-0.07}$ and $L^\ast\equiv \nu L_{\nu}^{\ast \, 8 \,\rm \mu m} = (1.32^{+1.31}_{-0.74})\times 10^{11} \, \rm L_\odot$.}
\end{figure}

 The shape of the $z\sim0$ LF can be fitted with a double-exponential function (Saunders et al.~1990; Pozzi et al.~2004; Le Floc'h et al.~2005):

\begin{eqnarray}
\label{eq-dexp}
\Phi(L) \, d\log_{10}(L)=\Phi^\ast \, \left(\frac{L}{L^\ast}\right)^{1-\alpha} \times && \nonumber\\
\times \,
\exp\left[ - \frac{1}{2\sigma^2} \, \log^2_{10}\left(1+\frac{L}{L^\ast}\right)\right]  \, d\log_{10}(L),&&
\end{eqnarray}

\noindent where, in this case, $L \equiv \nu L_{\nu}^{8 \,\rm \mu m}$.  The parameters $\alpha$ and $1/\sigma^2$ correspond to the slopes at the faint and the bright ends, respectively. $L^\ast$ is the characteristic $\nu L_{\nu}^{\ast \,8 \,\rm \mu m}$ luminosity where the transition between the faint and bright regimes occurs, and $\Phi^\ast$ is the normalization factor. Usually, the parameter $\alpha$ is fixed a priori, as the  faint-end of the LF is poorly constrained. Fixing $\alpha=1.2$ (e.g. Zheng et al.~2006) and using a $\chi^2$ minimization technique, we obtain that the best-fitting parameters for the LF at $z\sim0$ are $\sigma=0.36\pm0.01$, $L^\ast=(5.8\pm0.2)\times10^9 \, \rm L_\odot$ and  $\Phi^\ast=(5.7\pm0.1)\times10^{-3} \, \rm Mpc^{-3} \, dex^{-1}$. The resulting curve (dotted line in Figure \ref{fig_lf8z1}) produces an excellent fitting of the $1/V_{\rm max}$ LF at $z\sim0$.

 Assuming that the form given in eq.(\ref{eq-dexp}) is also suitable to describe the IR LF for star-forming galaxies at higher redshifts, we obtain a second independent calculation of the rest-frame \egm LF at redshift $z=1$ using the STY (Sandage, Tammann \& Yahil, 1979) maximum likelihood (ML) analysis. This is a parametric technique which assumes a given shape for the LF. No data binning is involved in the calculation. The advantage of the ML analysis over the $1/V_{\rm max}$ technique is that the former does not contain any assumption on a uniform spatial distribution of galaxies.  The corresponding likelihood estimator reads:

\begin{equation}
\mathcal L [s_k | (z_i, L_i)_{i=1,...,N}] = \prod_{i=1}^N \, \left[\frac{\Phi(s_k, L)}{\int_{\log_{10} (L_{0}^{\mathnormal i})}^{+ \infty}\Phi(s_k, L) \, d\log_{10}(L)}\right]^{w_i} \!\!\!\!\!\!,
\end{equation} 

\noindent where the product is made over the $i=1,...,N$ galaxies of the sample.  $\Phi(s_k, L)$ is the adopted form for the LF as a function of the luminosity $L$, and which depends on the  parameters $s_k$. $L_{0}^{\mathnormal i}$ is the minimum luminosity at which  the $i$-th galaxy would be observable, given its redshift $z_i$ and the flux limit of the survey. The weighting factors $w_i$ allow to take into account completeness corrections (Zucca, Pozzetti \& Zamorani~1994; Ilbert et al.~2005). By maximizing $\mathcal L$ (or, for simplicity, its logarithm), one can obtain the values of the parameters $s_k$ yielding the maximum likelihood. The normalization factor $\Phi^\ast$ is recovered after the maximization, by integrating the obtained  maximum likelihood LF without normalization in the  range of luminosities of the survey, and making it equal to the number density of observed galaxies. We note that the ML analysis provides a direct calculation of the LF (i.e. it does not constitute a fitting procedure as the $\chi^2$ minimization) and is completely independent of the LF obtained with the $1/V_{\rm max}$ technique.

  For the case of our rest-frame \egm LF at $z=1$, we apply the STY method using eq.(\ref{eq-dexp}) and fixing the slopes at the faint and bright ends to the same values as at $z\sim0$, i.e. $\alpha=1.2$ and $\sigma=0.36$, respectively.  In this case, we obtain that the value of the characteristic luminosity which maximizes the likelihood estimator is  $L^\ast\equiv \nu L_{\nu}^{\ast \, 8 \,\rm \mu m} = (3.55^{+0.52}_{-0.40})\times 10^{10} \, \rm L_\odot$ and the derived normalization factor is $\Phi^\ast=(3.95^{+0.50}_{-0.49})\times 10^{-3} \, \rm Mpc^{-3} \, dex^{-1}$. The error bars on $L^\ast$ have been computed considering $\Delta(\ln \mathcal L)=-0.5$ and the uncertainties derived from the Monte Carlo simulations. The degeneracies in parameter space given by $\Delta(\ln \mathcal L)=-0.5$ dominate the $L^\ast$ error budget. The  error bars on $\Phi^\ast$ have been derived using the extreme values of $L^\ast$ (i.e $L^\ast \pm$ its error). The resulting curve for the ML LF at $z=1$,  obtained with a double-exponential law with $\sigma=0.36$, is indicated with a solid line in the upper panel of Figure \ref{fig_lf8z1}.

  Another possibility is to allow the slope at the bright end ($1/\sigma^2$) to be a free parameter in the ML analysis. Doing so, we obtain that the ML is produced for: $\sigma=0.20^{+0.11}_{-0.07}$, $L^\ast\equiv \nu L_{\nu}^{\ast \, 8 \,\rm \mu m} = (1.10^{+0.99}_{-0.64})\times 10^{11} \, \rm L_\odot$ and the derived normalization is $\Phi^\ast=(2.54^{+0.60}_{-0.35})\times 10^{-3} \, \rm Mpc^{-3} \, dex^{-1}$ (dotted-dashed line in the upper panel of Figure \ref{fig_lf8z1}). The degeneracy in $(\sigma,L^\ast)$ space is shown in the lower panel of this same figure.

  The LF obtained with the ML analysis, both in the case of a free $\sigma$ value and fixed $\sigma=0.36$, are in good agreement with the LF computed  with the $1/V_{\rm max}$ method. This confirms  that the double-exponential law in eq. (\ref{eq-dexp}) also provides a good description of the \egm LF  at high redshifts. The degeneracy existing in the $\sigma$ value is due to the limited constraint that our data can put on the bright end of the LF at $z=1$. In Figure \ref{fig_lf8z1}, we see that the  
 double-exponential forms with $\sigma=0.20$ and 0.36  only differ significantly at the very bright end of the LF ($\nu L_{\nu}^{8 \,\rm \mu m} \gsim 10^{11.5} \, \rm L_\odot$ at $z=1$). Large-area surveys with a significant number of very bright IR galaxies, as for example the $\sim 2 \,\rm deg^2$ {\em Spitzer}-COSMOS survey (Sanders et al.~2006),  are necessary to  set tighter constraints in $(\sigma,L^\ast)$ space.

  Finally, we explore whether other functional forms could also be suitable to describe the rest-frame \egm LF at $z=1$.  We  repeat the calculation of  the LF with the STY method, but this time using a Schechter (1976) function:

\begin{equation}
\label{eq-sch}
\Phi(L) \, d\log_{10}(L)=\Phi^\ast \, \left(\frac{L}{L^\ast}\right)^{1-\alpha} \!\!\!\!\!\times \, \exp\left(-\frac{L}{L^\ast}\right) \, d\log_{10}(L).
\end{equation}

\noindent By fixing $\alpha=1.2$, we find that the maximum likelihood is obtained for a characteristic luminosity $L^\ast\equiv \nu L_{\nu}^{\ast \, 8 \,\rm \mu m} =(7.2^{+0.9}_{-0.7})\times 10^{10} \, \rm L_\odot$ and the derived normalization is  $\Phi^\ast=(3.88^{+0.46}_{-0.41})\times 10^{-3} \, \rm Mpc^{-3} \, dex^{-1}$. The resulting ML curve is shown with  a dashed line in Figure \ref{fig_lf8z1}. The Schechter form actually produces a LF quite close to that obtained with the $\sigma=0.20$ double-exponential form, within the observed luminosity range of our survey.  

  The degeneracy existing in the shape of the IR LF, as constrained from our data, produces some uncertainty in the determination of the number density of the most luminous IR galaxies (cf. Table \ref{tab_numd}). However, as we discuss below, this degeneracy has little impact on the derived luminosity density. This value is mainly governed by the turnover of the LF, which we can properly determine here, given the depth of our survey.

\subsection{The evolution of the rest-frame \egm LF from $z\sim0$ to $z=1$}   
\label{sec_evz1}

 When using the same law to describe the rest-frame \egm LF both at redshifts  $z\sim0$ and $z=1$, the differences found in the characteristic luminosity $L^\ast$ and the normalization parameter $\Phi^\ast$ can be understood as a luminosity and density evolution:

\begin{eqnarray}
\label{eq-ldev1}
L^\ast(z_2=1)=L^\ast(z_1\sim0) \times  \left(\frac{1+z_2}{1+z_1}\right)^{\rm \gamma_L} && \nonumber\\
\Phi^\ast (z_2=1)= \Phi^\ast(z_1\sim0) \times  \left(\frac{1+z_2}{1+z_1}\right)^{\rm \gamma_\delta},  &&
\end{eqnarray}

\noindent where we strictly use $z_1=0.2$ (the median redshift of the Huang et al. sample).  $\gamma_{\rm L}$ and $\gamma_\delta$ describe the evolution of the $L^\ast$ and $\Phi^\ast$ parameters with redshift. The values of  these parameters at $z\sim0$ and $z=1$, corresponding in both cases to a double-exponential with $\sigma=0.36$,  produce (cf. Section \ref{sec_mlz1}):

\begin{eqnarray}
\label{eq-gamma_01}
\rm \gamma_L &=& \,\,\,\,  3.5 \pm 0.4 \nonumber\\
\gamma_{\delta}&=& -0.7 \pm 0.1. 
\end{eqnarray}

\noindent This implies a strong positive luminosity evolution and a mild negative density evolution between $z\sim0$ and $z=1$.  The  mild negative density evolution to $z=1$ refers to the overall normalization  $\Phi^\ast$. However, it is clear from Figure \ref{fig_lf8z1} that, within the \egm luminosity range spanned by our sample, the density of galaxies at $z=1$ is larger than that at $z\sim0$. This is consistent to what has been found by Le Floc'h et al.~(2005) from the analysis of the rest-frame $15 \, \mu \rm m$ LF and confirms, once more, the increasing importance of IR galaxies up to redshift $z\sim1$. For clarity, the density of galaxies we obtain by integrating the rest-frame \egm LF above different luminosity cuts at different redshifts are shown in Table \ref{tab_numd}.

 By integrating the LF weighted by the luminosity values, over all luminosities, we can obtain the total rest-frame \egm luminosity density. In fact, for the obtention of the total luminosity density, the precise limits of integration are irrelevant, provided the turnover of the LF is completely contained within these limits. Moreover, the use of any of the different laws which are suitable to describe the LF (see Section \ref{sec_mlz1}) produces basically the same value for the luminosiy density, as all of them are in close agreement around the turnover.
  
 At $z=1$, we find that the total rest-frame \egm luminosity density is $(1.4\pm0.1)$, $(1.3\pm0.1)$ and $(1.4\pm0.1)\times 10^8 \, \rm L_\odot Mpc^{-3}$ for the cases of a double-exponential law with $\sigma=0.36$, 0.20 and a Schechter function, respectively. This is $\sim 4.0-4.3$ times the corresponding luminosity density at $z\sim0$.

\section{The rest-frame \egm LF at redshift $\lowercase{z}\sim2$}
\label{sec_lf8}

\subsection{The rest-frame \egm LF for star-forming galaxies at redshift $\lowercase{z}\sim2$}
\label{sec_lf8sf}

 The selection of \tfm galaxies at redshifts $z\sim2$ is particularly suitable to compute the rest-frame \egm LF. The IR SED of star-forming galaxies is characterised by the presence of PAH emission lines from rest-frame wavelengths $\lambda=3.3$ through $17 \, \rm \mu m$ (D\'esert, Boulanger \& Puget 1990). In particular, one of the main features in the SED is the PAH bump around 7.7 and $8.6 \, \rm \mu m$, responsible for a positive selection effect on galaxies at $z\sim1.9$ at \tfm (Caputi et al.~2006a). The light associated with this bump produces  a substantial fraction of the observed \tfm  output at $z\sim2$ (the remaining part mainly being produced by AGN). The study of the rest-frame \egm LF for star-forming galaxies gives direct information on the luminosity distribution of PAH emission in IR galaxies.  In particular at $z\sim2$, it should allow us to understand this PAH emission distribution when the Universe was only one fourth of its present age.

  We compute the rest-frame \egm luminosity ($\nu L_{\nu}^{8 \,\rm \mu m}$)  of each galaxy in a similar way as for those galaxies at $0.9<z<1.1$. In this case,  the width of the redshift bin we consider $1.7<z<2.3$ implies that the observed \tfm maps rest-frame wavelengths $7.2<\lambda_{\rm rf} < 8.9 \, \rm  \mu m$. As we have seen in  Section \ref{sec_kcz1}, the k-corrections are usually computed based on SED models, which have been calibrated using local IR galaxy templates and other observables.  However, we showed that this can be somewhat controversial, especially in  the PAH-dominated region, where different models show important discrepancies.  To compute the k-corrections from $7.2-8.9$ to \egm, we can avoid relying on any IR SED model by directly using measured rest-frame IR spectra of star-forming galaxies at redshifts $z \gsim 1.5$, convolved with the \tfm-filter transmission function.
 These spectra have been obtained with the Infrared Spectrograph (IRS) for {\em Spitzer} (Lutz et al.~2005; Yan et al.~2005).  These spectra correspond to star-forming ULIRG which are on average brighter than those studied here.  In spite of that, the k-corrections derived within the PAH region from these galaxies are expected to be applicable to our galaxies. For example, the equivalent widths of PAH lines in the Yan et al. (2005) star-forming galaxies are comparable to those of other lower luminosity ULIRG. In general,  PAH line equivalent widths appear to be quite independent of the bolometric IR luminosities of star-forming galaxies (Peeters et al.~2004; Brandl et al.~2006).

  For the wavelength range considered, the k-correction factors derived from  empirical spectra vary between $k=1$ (at $\lambda_{\rm rf}=8 \, \rm \mu m$) and $k=1.44 \pm 0.36$ (at $\lambda_{\rm rf}=8.9 \, \rm \mu m$). These k-corrections are in good agreement with those predicted by the Lagache et al.~(2004) models. The median of the differences is $\sim 5\%$ in the considered wavelength range ($7.2-8.9 \, \rm \mu m$). Thus, the use of  empirical k-corrections for our rest-frame \egm LF at $z\sim2$ is consistent with the use of model-dependent k-corrections at $z=1$.

\begin{figure} 
\epsscale{1.0}
\plotone{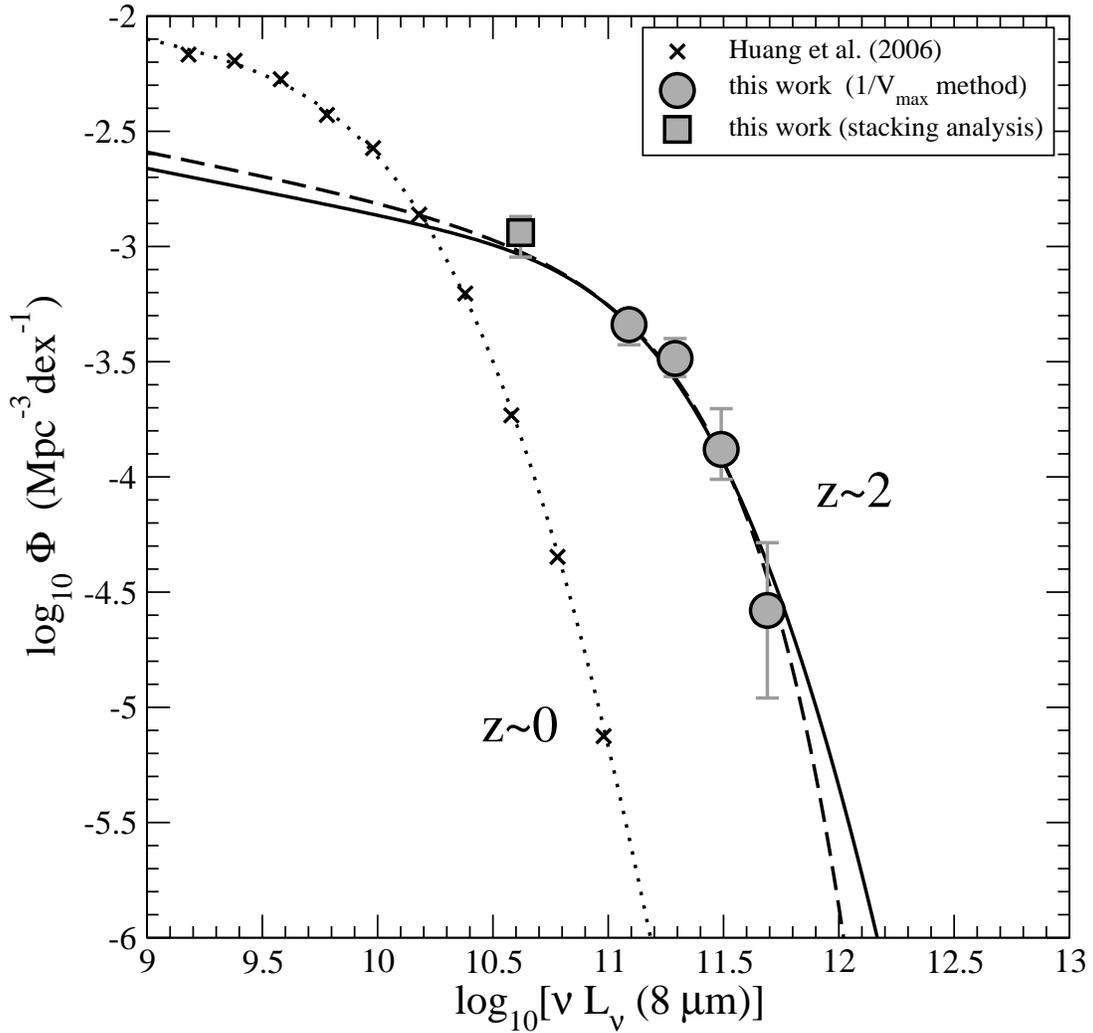}
\caption[]{\label{fig_lf8ml} The rest-frame \egm LF for star-forming galaxies at $z\sim2$ in the GOODS fields. The filled circles show the LF in the region of completeness of \egm luminosities, as computed with the $1/V_{\rm max}$ method.   The solid and dashed lines show the \egm LF at $z\sim2$ computed with the ML STY method, assuming a double-exponential form as in eq. (\ref{eq-dexp}) and a Schechter function, respectively.  The filled square is an extension of the LF at the faint-end, obtained using stacking analysis (see text for details).  The addition of this point {\em a posteriori} allows to validate the extrapolated shape of the LF at the faint end. The \egm LF at $z\sim0$ computed by Huang et al.~(2006) has also been added for a comparison.}
\end{figure}

 As at redshift $z=1$, we compute the rest-frame \egm LF at redshift $z\sim2$ alternatively using the $1/V_{\rm max}$ method and the ML analysis. For the star-forming galaxy LF at this redshift, we consider the 132 star-forming galaxies with $1.7<z<2.3$ within our sample. The rest-frame \egm LF for star-forming galaxies at $z\sim2$ computed with the two methods is shown in Figure \ref{fig_lf8ml} (filled circles for the $1/V_{\rm max}$ method and solid and dashed lines for the ML analysis). The total comoving volume probed at these redshifts is $5.7 \times 10^5 \, \rm Mpc^{-3}$. 
 
 For the $1/V_{\rm max}$ calculation, we computed the error bars taking into account Poisson statistics and the errors on photometric redshifts and k-corrections through Monte-Carlo simulations. We constructed 1000 mock catalogues, each one containing 132 galaxies, as the original catalogue. The redshift of each source has been allowed to randomly vary following a Gaussian distribution centred at $z_{\rm centre}=\mathnormal z-0.01$ and with a dispersion $\sigma_z=0.06 \, (1+z)$ (cf.  Section \ref{sec_finsamp}), where $z$ is the redshift of the source in the original catalogue. The redshift of those sources with spectroscopic determinations have been left unchanged. To include the uncertainties in the k-corrections, we computed the rest-frame \egm luminosity $\nu L_{\nu}^{8 \,\rm \mu m}$ of each galaxy in the mock catalogue allowing the corresponding k-correction to have a random value within the range of its error bar. Once more,  we recompute the LF with the $1/V_{\rm max}$ method for each of the mock catalogues, with exactly the same procedure as for the original catalogue. The distribution of the LF values in each  $\nu L_{\nu}^{8 \,\rm \mu m}$ bin   determine the error bars on our original $1/V_{\rm max}$ LF. 
 
  The LF computed with the $1/V_{\rm max}$ method which is shown in Figure \ref{fig_lf8ml}  exclusively corresponds to the region of $\nu L_{\nu}^{8 \,\rm \mu m}$ completeness ($\nu L_{\nu}^{8 \,\rm \mu m} \gsim 10^{11}\, \rm L_\odot$). This is essential to assure that our results are not affected by incompleteness effects. 
 
 Also at these redshifts, we analyse the field-to-field variations computing the rest-frame \egm LF in the GOODS/CDFS and GOODS/HDFN separately. The results are shown in the right-hand panel of Figure \ref{fig_lf8separa}. We see that, in spite of the sample variance, the two LF are still consistent within the error bars.

 We perform the ML analysis for the combined fields in the same way as for galaxies at $0.9<z<1.1$. Once more, we assume that the double-exponential form given by eq. (\ref{eq-dexp}) with fixed slopes $\alpha=1.2$ and $1/\sigma^2=1/(0.36)^2$ can be used to describe the rest-frame \egm LF for star-forming galaxies at $z\sim2$. In this case, the number of galaxies is not sufficient to allow us to leave the bright-end slope as a free parameter (i.e. the ML algorithm does not converge to reasonable values). Also, the adoption of the same $\sigma$  value as at $z\sim0$ is useful to directly compare the resulting values of $L^\ast$ and $\Phi^\ast$ at different redshifts. Applying the STY method with a double-exponential with $\sigma=0.36$ to our star-forming galaxies at $z\sim2$, we obtain that the value 
  of the characteristic luminosity which maximizes the ML estimator is  $L^\ast\equiv \nu L_{\nu}^{\ast \, 8 \,\rm \mu m} = (8.3^{+1.5}_{-1.1})\times 10^{10} \, L_\odot$ and the derived normalization factor is $\Phi^\ast=(9.0^{+2.1}_{-1.7})\times 10^{-4} \, \rm Mpc^{-3} \, dex^{-1}$.  The resulting curve for the ML LF at $z\sim2$ is indicated with a solid line in Figure \ref{fig_lf8ml}. Once more, the LF obtained with the ML analysis is in good agreement to that computed  with the $1/V_{\rm max}$ method, confirming that the double-exponential form in eq. (\ref{eq-dexp}) also provides a good description of the \egm LF also at redshift $z\sim2$.

 As for the LF at $z=1$, a Schechter function also appears to be an alternative  suitable law to describe the rest-frame \egm LF for star-forming galaxies at $z\sim2$ with the ML STY method (dashed line in Figure \ref{fig_lf8ml}).

\subsection{Testing the faint-end of the LF through stacking analysis}

 As we mentioned in Section \ref{sec_mlz1}, the faint end slope of the IR LF is not well constrained even at $z\sim0$, and the common procedure is to fix this slope to a given value. One could, however, put into question whether the fixed slope value we use here ($\alpha=1.2$) is realistic to describe the faint-end of the IR LF at different redshifts. In the analysis of the IR LF at redshifts $0\lsim z \lsim1.2$, Le Floc'h et al.~(2005) concluded that the slope at the faint-end could not be much steeper than 1.2, as otherwise the faint \tfm number counts would be overproduced. This result has been confirmed by Zheng et al.~(2006), using the stacking analysis at \tfm of a large sample of $0.1\lsim z \lsim 1$ galaxies. The stacking analysis technique allows to gain about an order of magnitude in the IR-flux detection limit (Dole et al.~2006a; Zheng et al.~2006).

 We do a similar stacking analysis using  the $K_s<21.5$ (Vega mag) galaxy sample presented in Caputi et al.~(2006c) for the GOODS/CDFS.  We stack at \tfm all those galaxies (except AGN) with redshifts $1.7<z<2.3$ which are below the completeness limit of the $\nu L_{\nu}^{8 \,\rm \mu m}$ luminosities (i.e. $\nu L_{\nu}^{8 \,\rm \mu m} \lsim 10^{11}\, \rm L_\odot$ at $z\sim2$). This includes, of course, all those $K_s<21.5$ mag galaxies at $1.7<z<2.3$ in the GOODS/CDFS which are not identified in the $S(\rm 24 \, \mu m)>80 \, \rm \mu Jy$ catalogue for the same field. We find that the average \tfm flux of these stacked sources is $S(\rm 24 \, \mu m)=(49.3\pm1.7) \, \rm \mu Jy$, which implies an average rest-frame \egm luminosity $\log_{10}(\nu L_{\nu}^{8 \,\rm \mu m})\approx 10.6$. To incorporate this stacking point in our differential LF expressed per dex unit, we need to estimate the flux --and thus the luminosity-- range covered by the stacking sample. Also, we need to apply a correction factor which accounts for the fact that the $K_s<21.5$ sample loses completeness in identifying \tfm galaxies below the $S(\rm 24 \, \mu m)=80 \, \rm \mu Jy$  limit. We perform both steps using the \tfm number counts obtained by Papovich et al.~(2004). These number counts are already corrected for incompleteness in the \tfm detections below the flux completeness limits of the Papovich et al. samples. From the distribution of these number counts with \tfm flux, we obtain that the average \tfm flux of our $S(\rm 24 \, \mu m)<80 \, \rm \mu Jy$ sample will be well-reproduced if the stacked  galaxies  span the flux range $30 \lsim S(\rm 24 \, \mu m)<80 \, \rm \mu Jy$.  On the other hand, from the total number counts within this flux range and ignoring the effects of sample variance, we can obtain  the average identification completeness produced by the $K_s<21.5$ sample. We estimate that the $K_s<21.5$ sample allows us to identify $\sim 79\%$ of the \tfm galaxies with $30 \lsim S(\rm 24 \, \mu m)<80 \, \rm \mu Jy$. The inverse of the completeness fraction gives us the correction factor for the LF in the stacking luminosity bin.  An intrinsic assumption here is that the identification completeness derived for this flux range is the same at all redshifts, so it can be applied to our $1.7<z<2.3$ sample. This assumption seems to be very plausible (compare the redshift distributions of the $83 \, \rm \mu Jy$-limited and total \tfm samples in Figure 3 of Caputi et al.~2006a and see also Dole et al.~2006b).

 The resulting stacking point is indicated with a square symbol in Figure \ref{fig_lf8ml}. We note that we only add this point  to our rest-frame \egm LF at $z\sim2$ {\em a posteriori}, and it does not play any role in the ML analysis. The good agreement between the stacking analysis point and the ML curve confirms that the value fixed for the faint-end slope of the \egm LF is adequate, and significantly larger slopes would not reproduce the average density of faint IR galaxies.

 We attempted to do a similar stacking analysis for sources at redshifts $0.9<z<1.1$, in order to test the faint-end of the rest-frame \egm LF at redshift $z=1$. However, the stacking at \tfm of $K_s<21.5$ ($Ks<20.5$ mag) galaxies which are below the \egm luminosity completeness limit at those redshifts produces an average source with flux $S(24 \, \rm \mu m)=16.6$ (25.4 $\rm \mu Jy$). Unfortunately, no information on \tfm number counts are available for or below such faint fluxes. This fact prevented us to obtain an extension of the rest-frame \egm LF at $z=1$ for faint luminosities.

 
\subsection{The evolution of the rest-frame \egm LF for star-forming galaxies from $z\sim0$ to $z\sim2$}

\begin{figure} 
\plotone{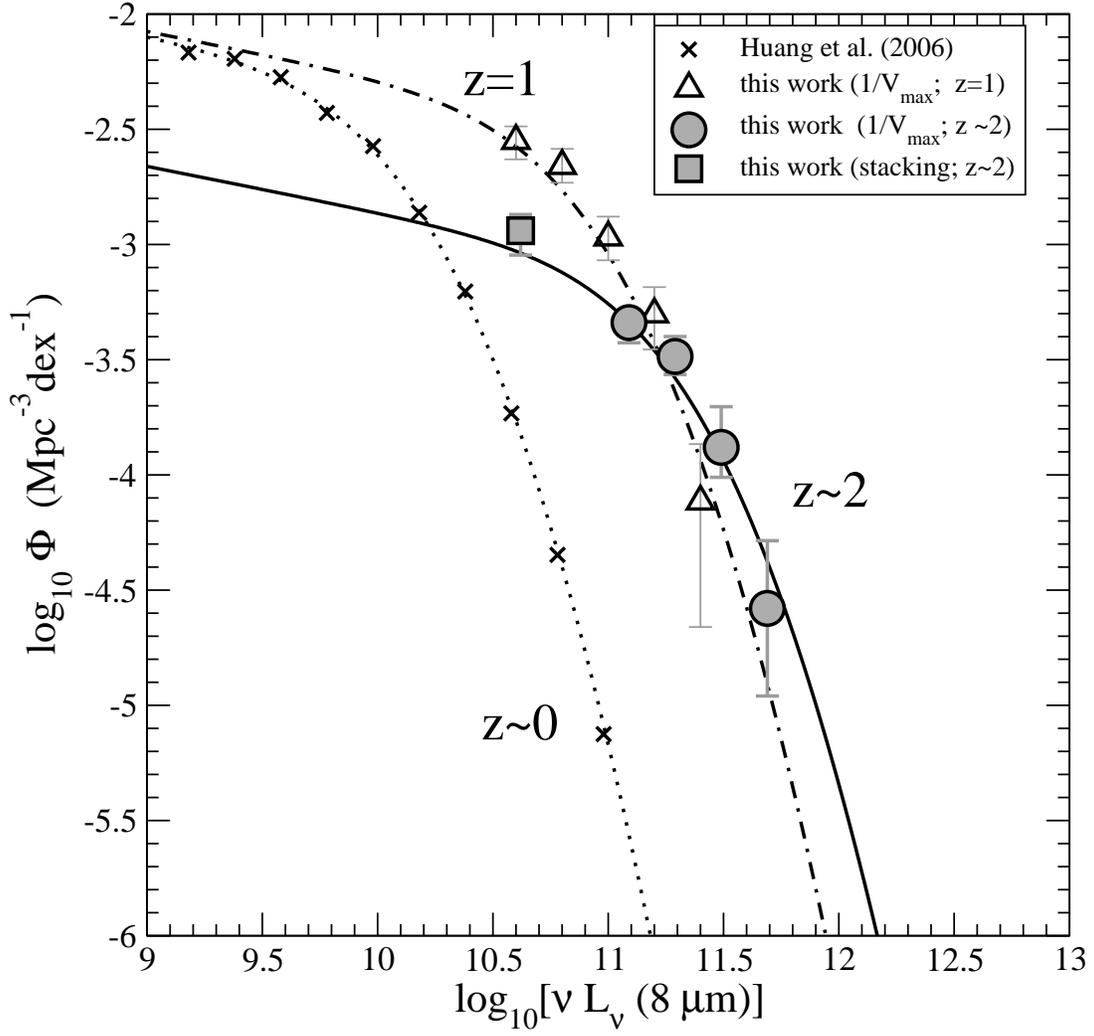}
\caption[]{\label{fig_lf8z1z2} The compared rest-frame \egm LF for star-forming galaxies at $z=1$ and $z\sim2$, both obtained in the GOODS fields. Symbols and lines are the same as in Figure \ref{fig_lf8ml}. The upward-pointing triangles correspond to the LF at $z=1$, as computed with the $1/V_{\rm max}$ method. The dotted-dashed line is the result of the ML analysis at the same redshift, adopting a double-exponential law with $\sigma=0.36$.}
\end{figure}

\begin{figure} 
\plotone{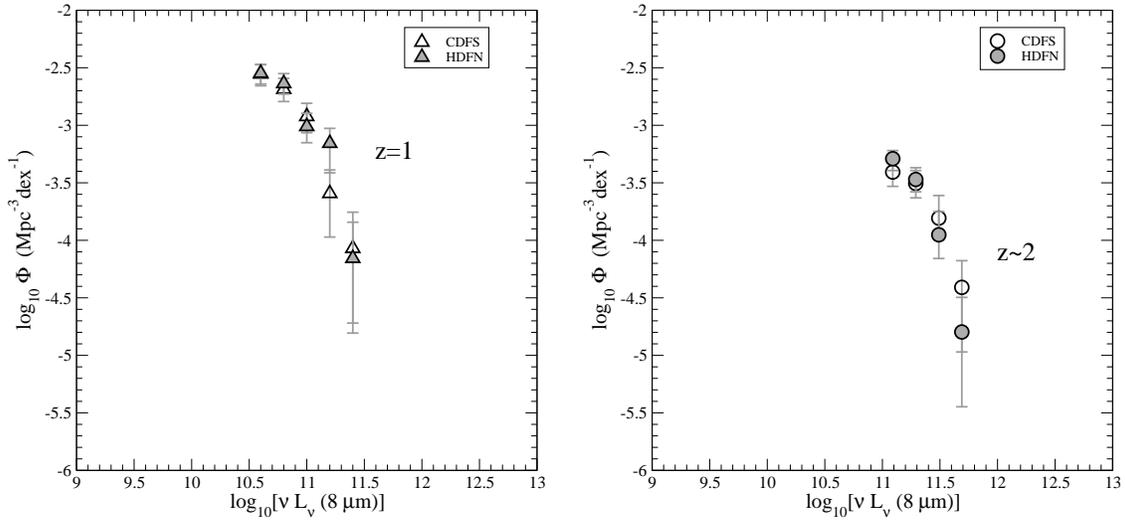}
\caption[]{\label{fig_lf8separa} The rest-frame \egm LF for star-forming galaxies in the GOODS/CDFS and HDFN separated, as computed with the $1/V_{\rm max}$ method. Left-hand panel: $z=1$; right-hand panel: $z\sim2$.}
\end{figure}

  We can now study the evolution of the rest-frame \egm LF from redshifts $z\sim0$ and $z=1$ to $z\sim2$. Figure \ref{fig_lf8z1z2} shows the three LF in a same plot. Different line styles in this plot correspond to a double-exponential form with $\sigma=0.36$. As in Section \ref{sec_evz1}, we can characterise the evolution of $L^\ast$ and $\Phi^\ast$ with redshift. If we consider

\begin{eqnarray}
\label{eq-ldev2}
L^\ast(z_2\sim2)=L^\ast(z_1\sim0) \times  \left(\frac{1+z_2}{1+z_1}\right)^{\rm \gamma_L} && \nonumber\\
\Phi^\ast (z_2\sim2)= \Phi^\ast(z_1\sim0) \times  \left(\frac{1+z_2}{1+z_1}\right)^{\rm \gamma_\delta},  &&
\end{eqnarray}

\noindent the derived $\rm \gamma_L$ and $\gamma_{\delta}$ at strictly $z_1=0.2$ and $z_2=1.93$ are
 
\begin{eqnarray}
\label{eq-gamma_02}
\rm \gamma_L &=& \,\,\,\,  3.0 \pm 0.4 \nonumber\\
\gamma_{\delta}&=& -2.1 \pm 0.4. 
\end{eqnarray}

 The obtained $\rm \gamma_L$ value indicates that the strong positive luminosity evolution of the rest-frame \egm LF continues up to redshift $z\sim2$. In contrast, the density evolution has quite a different trend between $z\sim0$ and $z=1$, and $z=1$ and $z\sim2$.  We showed in Section  \ref{sec_evz1} that the density of galaxies with $\nu L_{\nu}^{8 \,\rm \mu m} \gsim 10^{10.5} \, \rm L_\odot$ dramatically increases from $z\sim0$ and $z=1$. Between and $z=1$ and $z\sim2$, however, there appears to be a significant negative density evolution. If we write

\begin{eqnarray}
\label{eq-ldev3}
L^\ast(z_2\sim2)=L^\ast(z_1=1) \times \left(\frac{1+z_2}{1+z_1}\right)^{\rm \gamma_L} && \nonumber\\
\Phi^\ast (z_2\sim2)= \Phi^\ast(z_1=1) \times \left(\frac{1+z_2}{1+z_1}\right)^{\gamma_{\delta}},  &&
\end{eqnarray}

\noindent with strictly $z_1=1$ and $z_2=1.93$, we obtain 

\begin{eqnarray}
\label{eq-gamma_03}
\rm \gamma_L &=& \,\,\,\,  2.2 \pm 0.5 \nonumber\\
\gamma_{\delta}&=& -3.9 \pm 1.0. 
\end{eqnarray}

 A negative-density evolution at high ($z\gsim1$) redshifts have also been found with some of the fittings made for the $12 \, \mu \rm m$ LF by P\'erez-Gonz\'alez et al.~(2005). However, these authors conclude that the result of a negative density evolution should be taken with caution, as it could be produced by incompleteness in the faintest luminosity bins.  To test this, we repeat the ML analysis of our rest-frame \egm LF by considering only those galaxies with $S_\nu(24 \, \rm \mu m)> 120 \rm \mu Jy$ (which is roughly equivalent to excluding the faintest luminosity bin in the $1/V_{\rm max}$ method). In this case, the resulting normalization parameter value $\Phi^\ast$  implies $\gamma_\delta=-1.6 \pm 0.6$ and $\gamma_\delta=-2.7 \pm 1.3$ for the evolution between $z\sim0$ and $z\sim2$ and between $z=1$ and $z\sim2$, respectively. We conclude, then, that the negative density evolution result is not an effect of a plausible incompleteness at the faintest luminosities.

 It should be noted that all this analysis is based on the validity of the same law to describe the LF at different redshifts and the values obtained for  $\rm \gamma_L$ and   $\gamma_\delta$ depend on the adopted functional form. A more direct understanding of  the evolution of the rest-frame \egm LF can be achieved by comparing the integrated comoving number densities of galaxies above a given luminosity cut at different redshifts, as those we present in Table \ref{tab_numd}. If we restrict to the most luminous galaxies ($\log_{10}(\nu L_{\nu}^{8 \,\rm \mu m}) > 11.5$), we find that the number density increases with redshift up to $z\sim2$. For galaxies with $\log_{10}(\nu L_{\nu}^{8 \,\rm \mu m}) > 11$, remarkably, the number density appears to be basically the same at redshifts $z=1$ and $z\sim2$. Finally, if we consider those galaxies with $\log_{10}(\nu L_{\nu}^{8 \,\rm \mu m}) > 10.5$, we observe a clear change of trend between $z\sim0$ and $z=1$, and $z=1$ and $z\sim2$. While the number density of these galaxies increases by a factor $>20$ between $z\sim0$ and $z=1$, then the number density at $z=1$ decays to half its value by redshift $z\sim2$. We note that this decrement in intermediate-luminosity galaxies is not an effect of the faint-end slope $\alpha=1.2$ we assume for our LF. Inspection of Figure \ref{fig_lf8z1z2} shows that only a much higher $\alpha$ value (which would be inconsistent with the results of stacking analysis) could make equal the number densities of $\log_{10}(\nu L_{\nu}^{8 \,\rm \mu m}) > 10.5$ galaxies at $z=1$ and $z\sim2$.

 The rest-frame \egm luminosity density we derive at redshift $z\sim2$ is $7.5\pm0.5  \,(7.6\pm0.5)\times10^7 \, \rm L_\odot Mpc^{-3}$, obtained by integrating the  double-exponential (Schechter) function from the ML analysis. This represents more than twice the \egm luminosity density at $z\sim0$, but only half the corresponding luminosity density at $z=1$.

\subsection{The total rest-frame \egm LF at redshift $\lowercase{z}\sim2$}
\label{sec_total8}

 The rest-frame \egm LF at $z\sim2$ we presented in Section \ref{sec_lf8sf} has been calculated only taking into account the star-forming galaxies in our \tfm-selected sample at $1.7<z<2.3$. In this Section, we recompute the rest-frame \egm LF at $z\sim2$ for the GOODS fields considering all the 161 \tfm-selected star-forming galaxies and AGN with $1.7<z<2.3$. 
 
 We compute the rest-frame \egm luminosities as explained in Section \ref{sec_vmaxz1}. To determine the k-corrections for the AGN in our sample, we assume that their SED follow a power-law $f_\nu \propto \nu^{\alpha_{\rm SED}}$ (with $\alpha_{\rm SED}<0$).  For each AGN, we determine the value of $\alpha_{\rm SED}$ using its IRAC \egm and MIPS \tfm fluxes.

\begin{figure}
\plotone{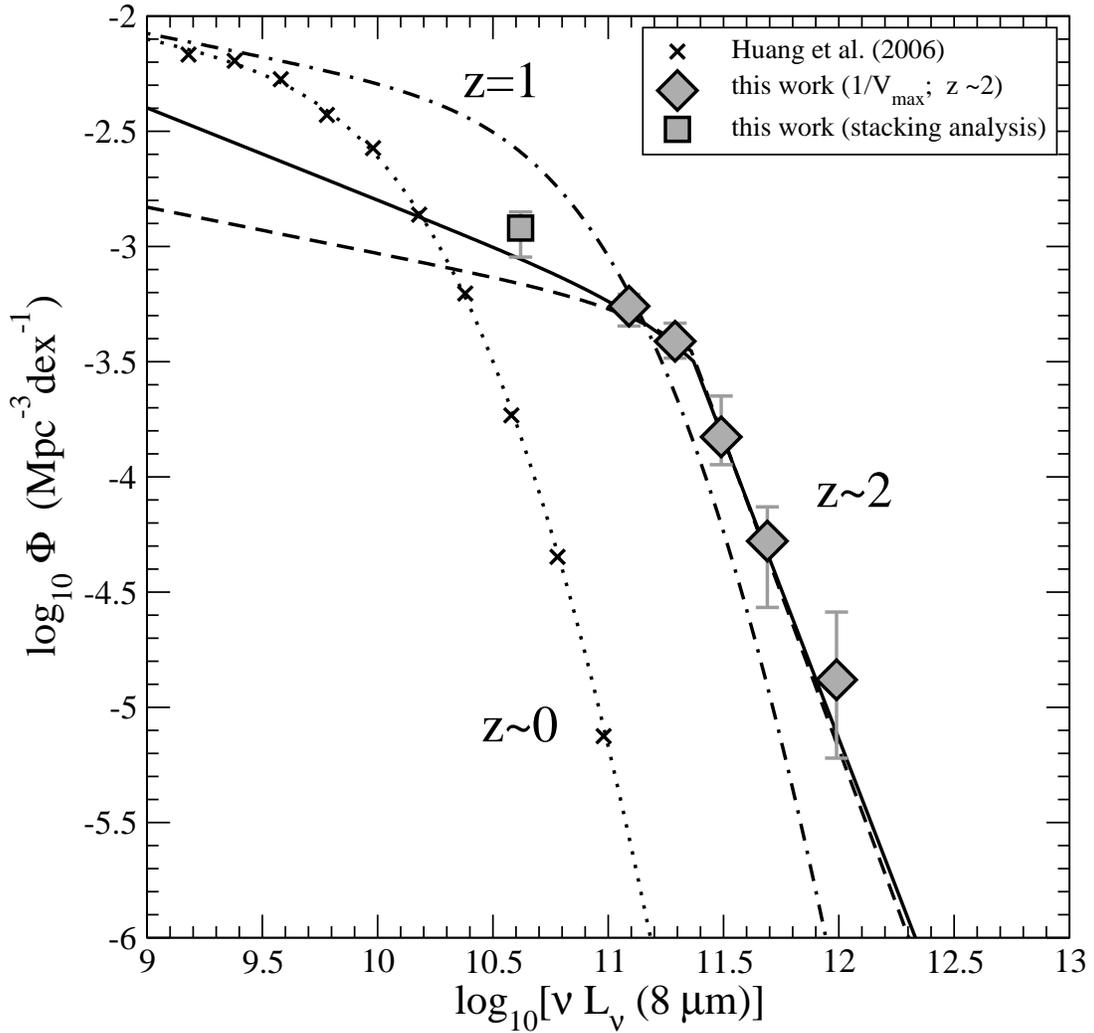}
\caption[]{\label{fig_lf8all} The rest-frame \egm LF for all the \tfm-selected galaxies (i.e. star-forming galaxies and AGN) at $z\sim2$. The diamond-like symbols indicate the LF computed with the $1/V_{\rm max}$ method. The  dashed and solid lines show the LF computed with the ML analysis, assuming the functional form given in eq.~(\ref{eq-exppl}) with $\alpha=1.2$ and 1.4, respectively. The remaining symbols and line styles are the same as in Figure \ref{fig_lf8z1z2}.}
\end{figure}

 The results of the total \egm LF calculated with the $1/V_{\rm max}$ method are indicated with filled diamonds in Figure \ref{fig_lf8all}. The error bars take into account Poisson errors and Monte Carlo simulations on the redshift and luminosity catalogues, as explained in Section \ref{sec_lf8sf}. Comparing this total \egm LF with that obtained  only for star-forming galaxies  (Figure \ref{fig_lf8ml}), we can see that AGN  mainly dominate the very bright end.  This excess of very bright sources suggests that neither the double-exponential form given in eq. (\ref{eq-dexp}) or a Schechter function are optimal to describe the bright end of the total \egm LF. At fainter magnitudes, however,the star-forming-galaxy and total LF show no significant difference, so we can safely assume the same behaviour at the faint-end.

 Thus, to compute the total rest-frame \egm LF with the STY method, we consider a combination of an exponential and a power-law, as follows:

\begin{equation}
\label{eq-exppl}
\Phi(L)=\left\{
\begin{array}{lrl}
    \Phi^\ast \, \frac{1}{const} \, \left(\frac{L}{L^\ast}\right)^{1-\alpha} \times  \\
   \times   \exp\left[ - \frac{1}{2\sigma^2} \, \log^2_{10}\left(1+\frac{L}{L^\ast}\right)\right], & \mbox{if $L \leq L^\ast$}& \\
 \\
    \Phi^\ast \, \left(\frac{L}{L^\ast}\right)^{1-\beta},  &\mbox{if $L > L^\ast$}&
  \end{array} \right.
\end{equation}

\noindent where $\beta$ is the slope at the bright end and the constant $const=\exp\{[-1/(2\sigma^2)] \,  \log_{10}^2 (2)\}$ guarantees continuity at $L=L^\ast$.  The stacking analysis point (square in Figure \ref{fig_lf8all}) is only added a posteriori to check the consistency of the results. In contrast to the \egm LF for star-forming galaxies,  the bright end of the total \egm LF  is sufficiently well constrained as to allow us to  leave $\beta$ as a free parameter.  At the faint-end, we fix $\alpha=1.2$ and $\sigma=0.36$, as in Section \ref{sec_lf8sf}.  The free-parameter values which yield the maximum likelihoood are: $\beta=3.7^{+0.4}_{-0.3}$, $L^\ast\equiv \nu L_{\nu}^{\ast \, 8 \,\rm \mu m} = (2.29^{+0.16}_{-0.15})\times 10^{11} \, \rm L_\odot$ and the derived normalization  is $\Phi^\ast=(3.52^{+0.16}_{-0.13})\times 10^{-4} \, \rm Mpc^{-3} \, dex^{-1}$. The resulting ML function is plotted with a dashed line in Figure \ref{fig_lf8all}. We observe that, while this ML LF is in very good agreement with that obtained from the $1/V_{\rm max}$ method, the stacking analysis point indicates that the faint-end is being under-produced.

  At luminosities $11.0 \lsim \log_{10}(\nu L_{\nu}^{8 \,\rm \mu m})\lsim 11.4$,  the  $1/V_{\rm max}$ \egm LF for star-forming and all galaxies are basically coincident. However, the slope value $\alpha$ which was suitable to describe the former does not seem sufficient to explain the faint-end of the total LF.   The explanation for this apparent contradiction is that the values of the different free parameters are coupled, and actually the definition of faint/bright ends depends on the value of $L^\ast$. In the case of the total rest-frame \egm LF at $z\sim2$, the value of the characteristic luminosity  $L^\ast$ is significantly higher than the ML value of $L^\ast$ for the star-forming galaxy LF. We recompute then the STY ML estimator for the total LF fixing the slope to a higher value $\alpha=1.4$.  The free-parameter values which yield the maximum likelihoood in this case are: $\beta=3.6^{+0.5}_{-0.3}$, $L^\ast\equiv \nu L_{\nu}^{\ast \, 8 \,\rm \mu m} = (2.34^{+0.29}_{-0.15})\times 10^{11} \, \rm L_\odot$ and the derived normalization  is $\Phi^\ast=(3.17^{+0.15}_{-0.28})\times 10^{-4} \, \rm Mpc^{-3} \, dex^{-1}$. The ML values of $\beta$ and $L^\ast$ are in agreement with those corresponding to $\alpha=1.2$, within the error bars. This indicates the robustness of the determination of the bright end and the turnover of the total \egm LF with our survey.  The resulting ML curve for the case with $\alpha=1.4$ is plotted with a solid line in Figure \ref{fig_lf8all}. This new curve appears to be consistent with the stacking analysis point.

 By integrating the obtained STY LF, we can compute the \egm luminosity density associated with the total IR galaxy population at $1.7<z<2.3$. This luminosity density is $\sim (9.0\pm0.6)\times 10^{7} \, \rm \L_\odot Mpc^{-3}$, i.e. $\sim 2.7$ times the total \egm luminosity density at $z\sim0$. Comparing the total \egm luminosity density at $z\sim2$ to that for only star-forming galaxies at the same redshift $(7.5\pm0.5) \times 10^7 \, \rm L_\odot Mpc^{-3}$, we conclude that AGN have a minor contribution  to this luminosity density even at high $z$ ($\sim 17\%$ at $z\sim2$).

\section{The bolometric IR LF at redshifts $\lowercase{z}=1$ and $\lowercase{z}\sim2$} 
\label{sec_lfbol}

\subsection{The conversion from \nLnegm to bolometric \lb}
 
\subsubsection{A new empirical calibration based on  Spitzer galaxies}

  In Section \ref{sec_lf8}, we studied the rest-frame \egm LF at redshift $z\sim2$ and its evolution from $z\sim0$. The aim of this section is to extend this study to the bolometric IR (i.e. $5 \lsim \lambda <1000 \, \rm \mu m$) LF.  The bolometric IR luminosity of a galaxy is produced by the thermal  emission of its gas content. In star-forming galaxies, the UV/optical radiation produced by young stars heats the  interstellar dust and the re-processed light is emitted in the IR. For this reason, in star-forming galaxies, the bolometric IR luminosity allows to obtain a direct and quite unbiased estimate of the current star-formation activity.

  Different methods to convert  $\nu \, L_\nu$ luminosities into bolometric IR luminosities \lb are common in the literature. Most of them rely on  calibrations made using nearby galaxies in {\em IRAS} or {\em ISO} catalogues (e.g. Chary \& Elbaz~2001; Elbaz et al.~2002; Takeuchi et al.~2005) or on the use of semi-empirical SEDs (e.g. Dale \& Helou 2002; Lagache et al.~2003, 2004; Dale et al. 2005). The extrapolation of these $\nu \, L_\nu$-\lb relations to high-redshift galaxies can be justified with different recent results. For example, Egami et al.~(2004) showed that composite SED of high-$z$ IR galaxies are well-described by local templates. Also, IR-galaxy models which assume such similarity in the SEDs can fit the 24, 70 and 160 $\rm \mu m$ number counts simultaneously (Lagache et al.~2004). Nevertheless, there is always some controversy on the large error bars which can be involved in  the $\nu \, L_\nu$-\lb conversions applied to high-redshifts. For example, Dale et al.~(2005) claim that the use of \tfm data (rest-frame \egm at $z\sim2$) can produce an uncertainty of up to a factor five in the derived bolometric IR luminosity of $z\sim2$ galaxies.

 To explore this issue, Bavouzet et al.~(2006) studied different $\nu \, L_\nu$-\lb relations using {\em Spitzer} low-to-intermediate redshift  galaxies.  Their sample consists of \tfm-selected galaxies with $R<20$ (Vega mag) in three different fields, namely the B\"{o}otes and the {\em Spitzer} First Look Survey fields, and the extended CDFS. The selection criterion of this sample is the detection of each galaxy in the IRAC \egm channel  and in all the three MIPS bands (i.e. at 24, 70 and 160 $\rm \mu m$).   All these galaxies have either spectroscopic or   COMBO17 photometric redshifts and span the redshift range $0.0\lsim z \lsim0.6$. AGN have been removed from their sample.

 To measure the bolometric IR luminosity \lb of each galaxy at redshift $z$, Bavouzet et al.~(2006) used  the  8 through $160 \, \rm \mu m$ fluxes. To integrate the corresponding empirical SED in each case, they summed up the areas below contiguous rectangles centred at rest-frame wavelengths  $8/(1+z), 24/(1+z), 70/(1+z)$ and $160 \, \rm \mu m/(1+\mathnormal z)$. At longer  wavelengths, they approximated the SED beyond  $\lambda > [160+(160-70)/2]/(1+z)=205 \, \rm \mu m/(1+\mathnormal z)$ with a triangle of slope $-4$. This slope is consistent with  the modified black-body emission  in the far-IR  produced by big grains of dust  thermalised at a temperature $T \sim 15-20 \rm K$ (Draine \& Lee~1984; Contursi et al.~2001). In fact, Bavouzet et al.~(2006) found that the use of any slope between $-3.5$ and $-4.5$ produced variations $<1\%$ on the computed bolometric luminosities. It is important to emphasize that the measurements of bolometric IR luminosities made by Bavouzet et al. are purely based on {\em Spitzer} data and are completely independent of any model template.

\begin{figure}
\plotone{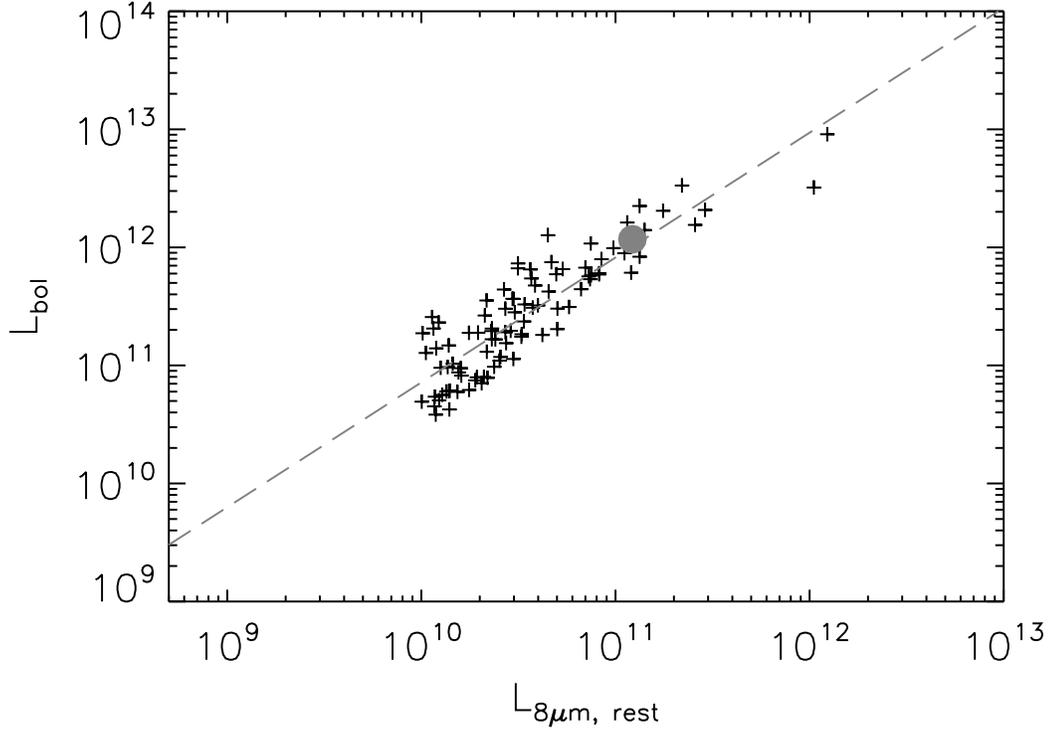}
 \caption[]{\label{fig_nicol8lir} The bolometric IR- versus rest-frame \egm-luminosity relation for galaxies with \nLnegm$> 10^{10} \, \rm L_\odot$ in the Bavouzet et al.~(2006) sample. The cross-like symbols indicate individual galaxies at redshifts $0.0<z<0.6$. The dashed line shows the best-fit relation. The filled circle  shows the resulting average value of  (\nLnegm; \lb) for a sample of galaxies at $1.3<z<2.3$, as obtained through stacking analysis in the GOODS/CDFS. This point shows that the average relation between \nLnegm and \lb  for  $1.3<z<2.3$ galaxies is basically the same as for galaxies at $0.0<z<0.6$.}
\end{figure}

 The resulting \lb versus rest-frame \nLnegm relation for the Bavouzet et al. sample  is shown in Figure \ref{fig_nicol8lir} (cross-like symbols). In this work, however, we restrict the analysis only to those 93 galaxies in the Bavouzet et al. sample  which have \nLnegm$> 10^{10} \, \rm L_\odot$ and signal-to-noise $S/N>3$ ratio in all the MIPS bands. The rest-frame \egm luminosities have been obtained applying k-corrections which do depend on different SED models (Chary \& Elbaz~2001; Elbaz et al.~2002; Lagache et al.~2004).  The   \nLnegm-\lb relation for these galaxies can be fitted with the following law (dashed line in Figure \ref{fig_nicol8lir}):
 
\begin{equation}
\label{eq-l8lir}
L_{\rm bol.}^{\rm IR}\,=1.91 \times [\nu L_\nu \rm (8 \, \mu m)]^{1.06}, 
\end{equation}  
 
\noindent with $\nu L_\nu \rm (8 \, \mu m)$ and $L_{\rm bol.}^{\rm IR}\,$ expressed in units of $\rm L_\odot$. The 1$\sigma$ dispersion for this relation is $\sim55\%$. This formula is directly applicable in all the redshift range $0.0 \lsim z \lsim 0.6$. We refer the reader to the Bavouzet et al.~(2006) paper for a generalised version of this formula including \nLnegm$< 10^{10} \, \rm L_\odot$ galaxies.

  To assess whether the formula displayed in eq. (\ref{eq-l8lir}) could also be  suitable to perform the  \nLnegm-\lb conversion for higher redshift galaxies, Bavouzet et al.~(2006) used the \tfm-selected galaxy samples in the GOODS/CDFS and HDFN (the same samples we use in this work). Of course, the bolometric luminosity of  the vast majority of $z \gsim 1$ galaxies cannot be empirically measured, as they are below the confusion limits of the {\em Spitzer}/MIPS images at 70 and $160 \, \rm \mu m$.  However, the average far-IR flux produced by  these high-redshift sources can be recovered through stacking analysis (Dole et al.~2006a). 
  
  Bavouzet et al.~(2006) stacked all those \tfm sources in the GOODS fields that lie at redshifts $1.3<z<2.3$,  with a median redshift $z \approx 1.68$.   The resulting (\nLnegm; \lb) value obtained with the stacking analysis is indicated with a filled circle in Figure \ref{fig_nicol8lir}. The \lb value for this point is corrected for the subestimation of the far-IR flux which is produced on high redshift sources by using the triangle-approximation method explained above.  This correction is about $10-15\%$. The locus occupied by the high-reshift stacked sources in the \nLnegm-\lb diagram strongly suggests that the relation described by eq. (\ref{eq-l8lir})  is also valid to link the \egm and bolometric IR luminosities of IR galaxies at $1.3<z<2.3$.

Thus, in this work we make use of the Bavouzet et al. relation described  by eq. (\ref{eq-l8lir}) to convert the rest-frame \egm of our star-forming galaxies into bolometric IR luminosities. We use these resulting bolometric IR luminosities to compute the corresponding LF for star-forming galaxies at $z=1$ and $z\sim2$ that we present in Section \ref{sec_lfir}.  As we explain in that section, the  55\% dispersion found for the \nLnegm-\lb relation largely dominates the error budget of the bolometric IR LF.

  As a final comment, we would like to discuss why the relation obtained by Bavouzet et al.~(2006) predicts a significantly smaller dispersion on the values of bolometric IR luminosities \lb obtained from rest-frame \egm fluxes than that predicted by Dale et al.~(2005). Firstly, the Dale et al.~(2005) sample includes  nearby galaxies of very different nature, and they even make separate studies of different IR regions within a same IR galaxy. Thus, because of its selection, it is expected that the Dale et al. sample displays a larger variety of IR properties than the Bavouzet et al.~(2006) sample.   Furthermore, to extrapolate their conclusions to high redshifts, Dale et al.~(2005) make use of the complete set of Dale \& Helou (2002) models. However, the majority ($\gsim 75\%$) of their wide range of observed SEDs only correspond to roughly half of these models (cf. figures in Dale et al.~2005).   The Bavouzet et al. sample has been selected  with a more homogeneous criterion and includes galaxies up to intermediate ($z\approx0.6$) redshifts. Thus, these galaxies are more likely representative of the typical  galaxies selected in IR surveys.  A thorough discussion of this issue is presented in  the Bavouzet et al.~(2006) paper.
  
\subsubsection{Comparison between different \nLnegm- \lb calibrations}
\label{sec_compl8lb}

 Several different laws to convert \nLnegm into bolometric IR luminosities \lb  are of common use in the literature. We analyse here how these different calibrations compare to the  relation empirically derived from {\em Spitzer} galaxies by Bavouzet et al.~(2006).

\begin{figure}
\plotone{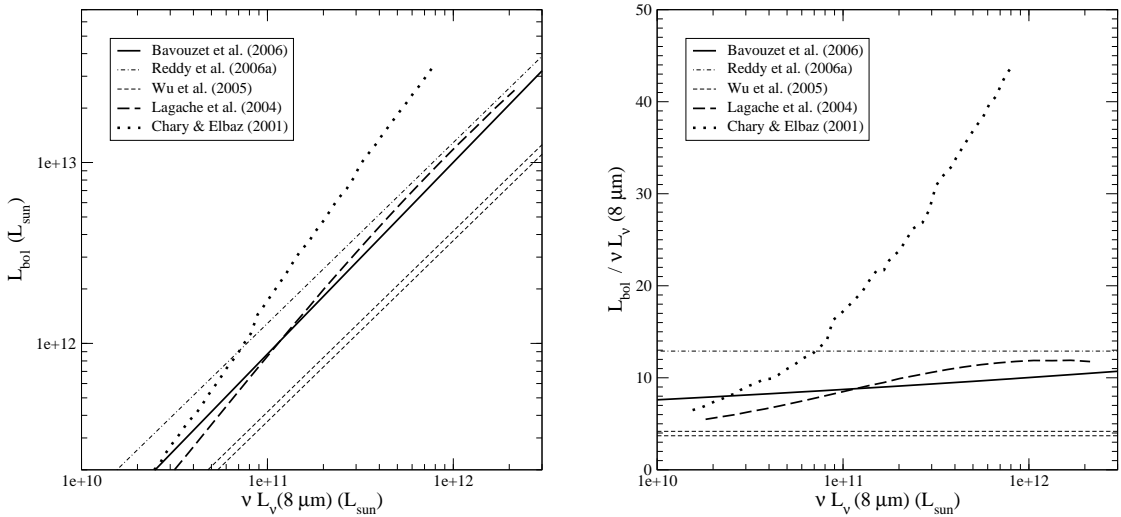}
 \caption[]{\label{fig_l8lbol} Comparison between different \lb versus \nLnegm relations  (left-hand panel) and  derived conversion factors versus \nLnegm (right-hand panel), as obtained from different calibrations available in the literature.}
\end{figure}

 Figure \ref{fig_l8lbol} shows the bolometric IR  \lb versus \nLnegm luminosity relations (left-hand panel) and the derived conversion factors \lb$/$\nLnegm (right-hand panel), both versus \nLnegm, as obtained using  different calibrations or derived from different IR SED models. The thick solid line shows the empirical relation obtained by Bavouzet et al.~(2006). The thick dashed and dotted lines correspond to the relations derived using the Lagache et al.~(2004) and the Chary \& Elbaz (2001) - Elbaz et al.~(2002) templates, respectively. To obtain these relations, we convolve all these templates with the transmission function of the MIPS \tfm filter. We find that the Lagache et al. model predicts a  \nLnegm- \lb relation which is in close agreement with the Bavouzet et al.~(2006) empirical calibration over all \egm luminosities. The Chary \& Elbaz templates, on the contrary, appear to over-produce the \nLnegm- \lb conversion. The differences with the Bavouzet et al. calibration are only within a factor $\sim2$ for galaxies with $\nu L_{\nu}^{8 \,\rm \mu m} < 10^{11} \, \rm L_\odot$, but become dramatically larger at higher luminosities.
 
 Previous comparisons of the $\nu \, L_\nu$-\lb relations predicted by different models have not detected such dramatic differences (see e.g. Le Floc'h et al.~2005). These previous comparisons analysed longer rest-frame wavelengths, beyond the PAH-dominated region in the SEDs. The comparison we present here is made in the most critic SED region, where different models show the largest discrepancies (cf. also Figure \ref{fig_12to8k}).  From this comparison, we find that the use of the Chary \& Elbaz templates to convert \nLnegm into \lb luminosities leads to significantly over-produced  bolometric IR  luminosity values for galaxies with $\nu L_{\nu}^{8 \,\rm \mu m} > 10^{11} \, \rm L_\odot$. 
 
 In Figure \ref{fig_l8lbol}, we also show the \nLnegm to \lb derived from the Wu et al.~(2005) formulae (thin dashed lines), which link \egm luminosities and star formation rates. The bolometric IR luminosities have been recovered using $SFR=1.72\times10^{-10} \, L_{\rm IR}$ (Kennicutt 1998).   Finally, the thin dotted-dashed line shows the relation used in Reddy et al.~(2006a). In this latter relation,  the \egm luminosities refer to the convolution in the wavelength range $\sim 5-8.5 \,\rm \mu m$, which is somewhat different from the transmission windows of the MIPS \tfm filter  ($\sim 6.6-9.4 \, \rm \mu m$ at $z\sim2$) or the IRAC \egm filter ($\sim 6.5-9.5 \, \rm \mu m$; Fazio et al. 2004). Once corrected for this difference, the Reddy et al.~(2006a) relation  becomes closer to the Bavouzet et al.~(2006) {\em Spitzer} calibration.

 In this work use the new {\em Spitzer}-based calibration given by eq.~(\ref{eq-l8lir}) to convert \nLnegm luminosities into bolometric IR luminosities \lb.  After computing the bolometric IR LF, we  analyse the contribution of LIRG and ULIRG to the total number and luminosity densities of IR galaxies at different redshifts. We warn the reader, however, on the implications of the differences between the \nLnegm- \lb  conversions shown in Figure \ref{fig_l8lbol}. For example, the Chary \& Elbaz conversion classifies as ULIRG to those sources with 
\nLnegm$\gsim 8\times 10^{10} \, \rm L_\odot$, while the Bavouzet et al. relation implies that only galaxies with \nLnegm$\gsim 1.1-1.2\times 10^{11} \, \rm L_\odot$ are ULIRG. These differences should be kept in mind when comparing different results from the literature, where different conversion laws are used.

\subsection{The bolometric IR LF for star-forming galaxies and its evolution to redshift $\lowercase{z}\sim2$}
\label{sec_lfir}

\begin{figure} 
\plotone{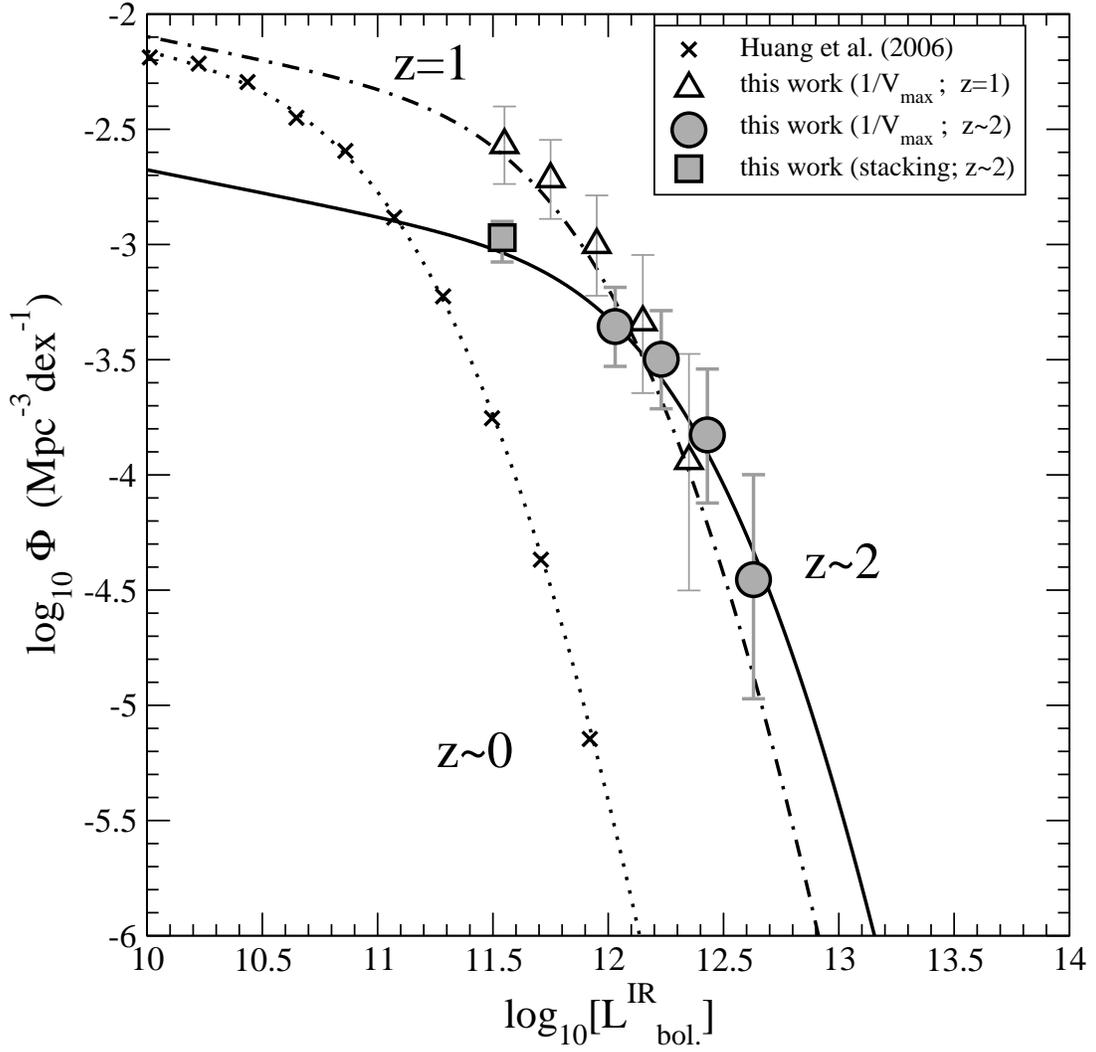}
\caption[]{\label{fig_lbolev} The evolution of the bolometric IR LF for star-forming galaxies from redshift $z=0$ to $z\sim2$. Symbols and line styles are the same as in Figure \ref{fig_lf8z1z2}.}
\end{figure}

 As we have seen in Section \ref{sec_compl8lb}, some calibrations used in the literature to convert \egm into bolometric IR luminosities  are quite discrepant with the empirical calibration obtained from {\em Spitzer} galaxies. Thus, to properly compare the bolometric IR LF at different redshifts, we convert the different \egm LF using the Bavouzet et al.~(2006) relation shown in eq. (\ref{eq-l8lir}). The results are shown in Figure \ref{fig_lbolev}.

 Firstly, we transform the Huang et al.~(2006) \egm LF at $z\sim0$ and compute the corresponding minimum $\chi^2$-fitting, using the functional form given in eq. (\ref{eq-dexp}). For the bolometric IR LF at $z\sim0$, we obtain the following best-fit parameter values: $\sigma=0.39 \pm 0.01$, $L^\ast_{\rm IR}=(4.0\pm0.2)\times10^{10} \, \rm L_\odot$ and  $\Phi^\ast=(5.4\pm0.1)\times10^{-3} \, \rm Mpc^{-3} \, dex^{-1}$.  The resulting best-fit curve to the bolometric IR LF at $z\sim0$ is shown  with a dotted line in Figure \ref{fig_lbolev}.

 The best-fit value we find for the slope at the bright-end at $z\sim0$, i.e. $\sigma=0.39$, is very similar to the value obtained for the bright-end slope of the rest-frame \egm LF ($\sigma=0.36$) at the same redshift.  This similarity is due to the fact that the \nLnegm-\lb conversion is quasi-linear.

 At redshifts $z=1$ and $z\sim2$, we compute the bolometric IR luminosities \lb of all our star-forming galaxies in the relevant redshift ranges by transforming their rest-frame \egm luminosities \nLnegm using eq. (\ref{eq-dexp}). We then obtain the bolometric IR LF using, alternatively, the $1/V_{\rm max}$ method and the ML STY analysis.

 The upward-pointing triangles and circles in Figure \ref{fig_lbolev} show the bolometric IR LF at $z=1$ and $z\sim2$, respectively, both computed with the $1/V_{\rm max}$ method.  These LF are only shown in the bins of completeness in \lb luminosities, given the flux limits of our sample and the redshifts corresponding to each case.  As for the rest-frame \egm LF, we applied small correction factors to account for the $5-6\%$ identification incompleteness of the $S(\rm 24 \, \mu m)>80 \, \rm \mu Jy$ galaxy sample.  For both LF, the error bars have been determined through Monte Carlo simulations, in a similar way as described in Section \ref{sec_lf8sf}. The mock catalogues generated in the simulations  take into account the error bars in the redshift determinations, in the case of photometric redshifts. However, in the case of the bolometric luminosities, the error budget is mainly dominated by the uncertainty associated with the \nLnegm-\lb conversion. To  take into account this error, we assign to each galaxy in the mock catalogues a random bolometric IR luminosity. This random luminosity \lb  is taken from a Gaussian distribution centred at the  value given by eq.  (\ref{eq-l8lir}) for the corresponding galaxy and with a 55\% dispersion. The re-computation of the LF with the $1/V_{\rm max}$ method on all the mock catalogues allows us to determine the error bars on the original LF calculation.

 The dotted-dashed and solid lines in Figure \ref{fig_lbolev} indicate the bolometric IR LF at $z=1$ and $z\sim2$, respectively, obtained with the ML analysis. We computed the bolometric IR LF using the STY method, assuming the functional form described in eq. (\ref{eq-dexp}). The faint and bright-end slope values have been fixed to the $z\sim0$ values, i.e. $\alpha=1.2$ and $\sigma=0.39$, respectively. At $z=1$, we obtain that the  value of the characteristic luminosity which yields the ML is $L^\ast_{\rm IR}=(2.5^{+0.4}_{-0.3}) \times 10^{11} \, \rm L_\odot$. The corresponding normalization factor is $\Phi^\ast=(4.0^{+0.6}_{-0.5}) \times 10^{-3} \,  \rm Mpc^{-3} \, dex^{-1}$. At $z\sim2$, the ML characteristic luminosity is $L^\ast_{\rm IR}=(6.3^{+1.1}_{-0.9}) \times 10^{11} \, \rm L_\odot$ and the corresponding normalization factor is $\Phi^\ast=(9.2^{+2.2}_{-1.7}) \times 10^{-4} \,  \rm Mpc^{-3} \, dex^{-1}$ (cf. Table \ref{tab_bol}). The error bars  on $L^\ast_{\rm IR}$ include the uncertainty produced by the 55\% dispersion in the \nLnegm-\lb relation, incorporated through the mock catalogues described above. Consistently with the results obtained in Sections \ref{sec_mlz1} and \ref{sec_lf8sf}, the LF independently  calculated with the $1/V_{\rm max}$ method and the ML STY technique are in very good agreement.

 Using also the \nLnegm-\lb relation given in eq. (\ref{eq-l8lir}), we compute the corresponding contribution of the stacked galaxies at $z\sim2$ which are below the \lb completeness limit of the sample, to the bolometric IR LF. Once more, the stacking analysis point appears in very good agreement with the extrapolation given by the ML analysis at the faint-end of the LF.

 Given the quasi-linearity of the \nLnegm-\lb conversion, the evolution we find for the bolometric IR LF from $z\sim0$ to $z\sim2$ is very similar to the evolution observed for the rest-frame \egm LF. For the bolometric IR LF, this implies:

\begin{itemize}
 
\item The number density of galaxies with $L_{\rm bol.}^{\rm IR} \gsim 10^{11} \, \rm L_\odot$ substantially increases from the local Universe to $z=1$ (cf. Table \ref{tab_bolnd}). This confirms the increasing importance of the LIRG and ULIRG populations between these redshifts (cf. e.g. Le Floc'h et al.~2005). 

\item Surprisingly, at $z\sim2$, the number density of star-forming ULIRG (i.e. sources with $L_{\rm bol.}^{\rm IR} > 10^{12} \, \rm L_\odot$) is only slightly larger than at $z=1$. This result is the combination of several factors: firstly, the exclusion of AGN in this analysis produces a relatively low density of ULIRG at $z\sim2$, as we have seen in Section \ref{sec_total8} that AGN dominate the bright end of the IR LF. Secondly, the use of the \nLnegm-to-\lb conversion given in eq. (\ref{eq-l8lir}), which, in comparison to the Chary \& Elbaz templates that are of common use in the literature, produces ULIRG only from larger  \nLnegm luminosities (cf. Figure \ref{fig_l8lbol}).

\item The number density of LIRG (i.e. sources with $10^{11}<L_{\rm bol.}^{\rm IR} < 10^{12} \, \rm L_\odot$) appears to be smaller at $z\sim2$ than at $z=1$. Although the limits of our survey do not allow us to directly observe LIRG at $z\sim2$,  the ML analysis suggests this results, which is in turn validated through the stacking analysis of $z\sim2$ $K_s$-band galaxies.
 
\end{itemize}

 Thus,  the ratio between the number densities of star-forming ULIRG and LIRG increases from $z=1$ to $z\sim2$. However, within our sample and given our star-forming galaxy/AGN separation, this effect appears to be mainly produced by a decrement in the density of LIRG by $z\sim2$, rather than a significant increment in the density of star-forming ULIRG. 
If our AGN separation criterion were excluding galaxies whose bolometric IR emission is actually dominated by star-formation, then the relative importance in the number density of star-forming ULIRG would be, of course, even larger at $z\sim2$. 

 We note that the decrement we find in the number density of LIRG between $z=1$ and $z\sim2$ is not influenced at all by the AGN separation criterion.
 
\subsection{Comparison with other works}

As we have seen in Section \ref{sec_compl8lb}, many different recipes are used in the literature to convert  $\nu \, L_\nu$ into \lb luminosities.  And even different conversions made from a same wavelength (in particular, rest-frame \egm) may lead to non-negligible discrepancies in the derived \lb luminosities. In spite of these differences, it is still instructive to compare the results of different bolometric IR LF calculations.

\begin{figure} 
\plotone{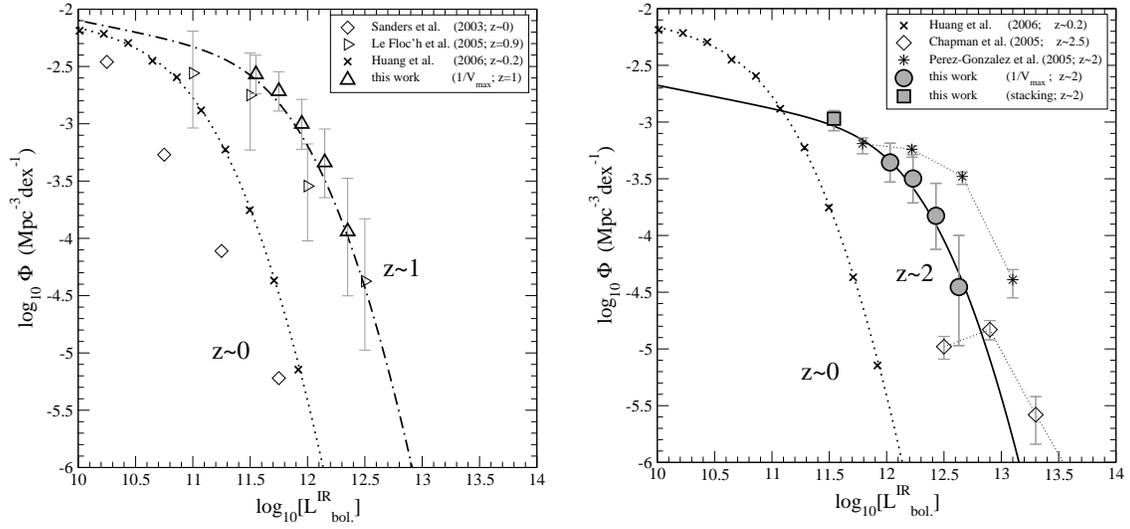}
\caption[]{\label{fig_lboloth} The  bolometric IR  LF obtained in this work compared to the determinations of other authors at similar redshifts: $z\sim1$ (left-hand panel) and $z\sim2$ (right-hand panel).}
\end{figure}

Figure \ref{fig_lboloth} compares the bolometric IR LF obtained in this work with those derived by other authors, at different redshifts. In the left-hand panel, we show the local bolometric IR LF computed from the {\em IRAS} revised galaxy sample (Sanders et al.~2003; diamond-like symbols) and  the bolometric IR LF  derived in this work from the Huang et al.~(2006) rest-frame \egm LF at $z\sim0.2$. The difference between the two is mainly due to a real evolution between $z=0$ and $z\sim0.2$. In the same panel, we also compare our bolometric IR LF at $z=1$ with that obtained by Le Floc'h et al.~(2005) at $z=0.9$. We observe that both LF are in good agreement, taking into account the error bars and the evolution expected between these redshifts  (cf. Le Floc'h et al.~2005).

In the right-hand panel of Figure \ref{fig_lboloth}, we show our bolometric IR LF at redshift $z\sim2$, compared to that derived from P\'erez-Gonz\'alez et al.~(2005) at a similar redshift and that computed from radio-detected sub-millimetre galaxies at $z\sim2.5$ (Chapman et al.~2005).

The bolometric IR LF derived from P\'erez-Gonz\'alez et al.~(2005; asterisks in Figure \ref{fig_lboloth}) has been obtained by converting their rest-frame $12 \, \rm \mu m$ LF at $z\sim2$, using the same recipe adopted by these authors to obtain bolometric IR luminosity densities (cf. eq.(1) in their paper). This conversion corresponds to the Chary \& Elbaz~(2001) $\nu L_\nu^{\rm 12 \, \mu m}$-\lb formula. Our bolometric IR LF at $z\sim2$ is in agreement, within the error bars, with that derived from  P\'erez-Gonz\'alez et al.~(2005) at luminosities $L_{\rm bol.}^{\rm IR} \lsim 10^{12.5} \, \rm L_\odot$. At brighter luminosities, however, the two LF present significant discrepancies. The differences between the two are produced by two factors: 1) the AGN exclusion:  P\'erez-Gonz\'alez et al.~(2005) only exclude the most extreme cases of AGN, while here we adopt a more extensive separation criterion; 2) the different $\nu \, L_\nu$-\lb conversions: as we have seen in Section \ref{sec_compl8lb}, the most drastic differences between the empirical {\em Spitzer}-based conversion we use in this work and that derived from the  Chary \& Elbaz~(2001) and Elbaz et al.~(2002) templates occur at luminosities $L_{\rm bol.}^{\rm IR} \gsim 10^{12} \, \rm L_\odot$.
This comparison illustrates the impact of using different $\nu \, L_\nu$-\lb relations, especially at high redshifts, where the most luminous IR galaxies are dominant.

The bolometric IR luminosities  derived from radio-detected sub-millimetre galaxies only can trace the very bright-end of the bolometric IR LF.  The  diamond-like symbols in the right panel of Figure \ref{fig_lboloth} correspond to the sub-millimetre-derived bolometric IR LF at $z\sim2.5$, as obtained by Chapman et al.~(2005). This LF does not exclude AGN and quickly loses completeness at $L_{\rm bol.}^{\rm IR} \lsim 10^{13} \, \rm L_\odot$. Taking into account these facts and the differences in redshift, we find that the Chapman et al.~(2005) bolometric IR LF at $z\sim2.5$ is consistent with our LF based on \tfm-selected galaxies at $z\sim2$.

\subsection{The evolution of the bolometric IR luminosity density}
\label{sec_ldens}

One of the final aims of computing the bolometric IR LF is to obtain an estimate of the IR luminosity density (in our case associated with star-forming galaxies) at a given look-back time. Previous works agree in a strong evolution of the IR luminosity density from the local Universe up to redshift $z\sim1$ (e.g. Flores et al.~1999; Gispert, Lagache \& Puget~2000; Pozzi et al.~2004; Le Floc'h et al.~2005). At higher redshifts, the situation is less clear as only recently IR facilities are allowing to put constraints on the IR Universe at $z\gsim1$.

 Given the discrepancies existing between different recipes to obtain bolometric IR luminosities (cf. Section \ref{sec_compl8lb}), we need to use the bolometric IR LF obtained with the same conversion at different redshifts, in order to properly compute the evolution of the IR luminosity density.

\begin{figure}
\includegraphics[width=12cm,angle=270]{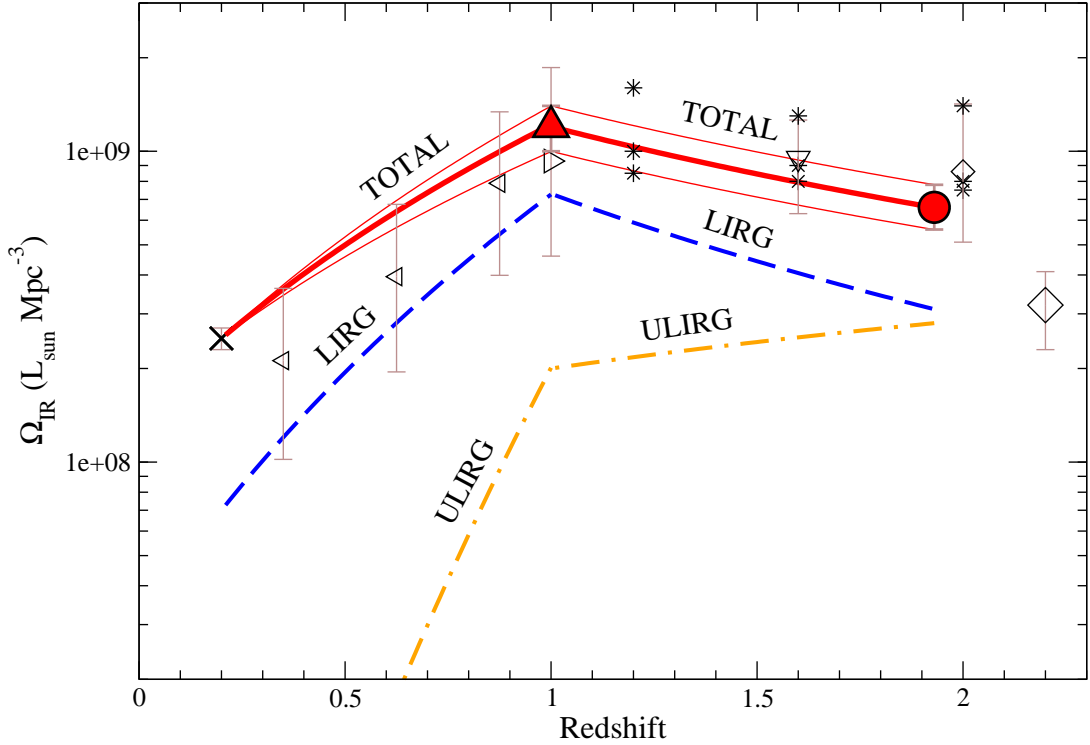}
\caption[]{\label{fig_lirdens} The  evolution of the comoving bolometric IR luminosity density with redshift.  The filled upward-pointing triangle and circle at redshifts $z=1$ and $z=1.93$ indicate the estimations of the respective  bolometric IR luminosity density obtained in this work: $\Omega_{\rm IR}=(1.2\pm0.2)\times 10^9$ and $(6.6^{+1.2}_{-1.0})\times 10^8 \, \rm L_\odot Mpc^{-3}$. The density at $z=0.2$ has been obtained from the bolometric IR LF  derived from the \egm LF by Huang et al.~(2006). The red thick solid line corresponds to an interpolation between these redshifts, assuming a $[(1+z_2)/(1+z_1)]^x$ evolution. The red thin solid lines indicate error bars on this evolution. Blue dashed and and orange dotted-dashed lines show the contributions of LIRG and ULIRG, respectively, at different redshifts. Other symbols refer to IR luminosity densities taken from the literature, and based on different datasets: {\em ISO} mid-IR  (Flores et al.~1999; left-hand-pointing triangles), {\em Spitzer} mid-IR (Le Floc'h et al.~2005 and P\'erez-Gonz\'alez et al.~2005; right-hand-pointing triangle and asterisks, respectively),  sub-millimetre (Barger et al.~2000 and Chapman et al.~2005; small and large diamond-like symbols, respectively) and radio (Haarsma et al.~2000; downward-pointing triangle). Some of these  IR luminosity densities have been obtained from the star-formation rate densities compiled by Hopkins et al.~(2004) and converted with the Kennicutt (1998) formula $SFR=1.72\times10^{-10} \, L_{\rm IR}$. }
\end{figure}

 Figure \ref{fig_lirdens} shows the evolution of the comoving IR luminosity density as a function of redshift.  Our determinations of the IR luminosity density at $z=1$ and $z\sim2$ (strictly $z=1.93$) are indicated with a filled upward-pointing triangle and circle, respectively: $\Omega_{\rm IR}(z=1)=(1.2\pm0.2)\times 10^9 \, \rm L_\odot Mpc^{-3}$ and $\Omega_{\rm IR}(z\sim2)=(6.6^{+1.2}_{-1.1})\times 10^8 \, \rm L_\odot Mpc^{-3}$. We obtain the values of these luminosity densites by integrating our respective bolometric IR LF obtained with the ML likelihood analysis, weighted with the luminosity values.  The error bars are determined by  the extreme cases of LF produced by the error bars on $L^\ast_{\rm IR}$.

 The cross-like symbol in Figure \ref{fig_lirdens} represents the bolometric IR luminosity density at $z\sim0$ (strictly $z=0.2$), as obtained from the bolometric IR LF derived from the Huang et al.~(2006) \egm LF: $\Omega_{\rm IR}(z\sim0)=(2.5\pm0.2)\times 10^8 \, \rm L_\odot Mpc^{-3}$.

 The thick solid line in Figure \ref{fig_lirdens}  interpolates the evolution of the total bolometric IR luminosity density between redshifts $z_1\sim0$ to $z_2=1$ and $z_1=1$ to $z_2\sim2$, assuming this evolution follows a $[(1+z_2)/(1+z_1)]^x$ law. Between redshifts $z\sim0$ and $z=1$, we find that the total bolometric IR luminosity density increases as  $[(1+z_2)/(1+z_1)]^{3.1\pm0.3}$ (where $z_1=0.2$ and $z_2=1.0$). This evolution is somewhat slower than that obtained by Le Floc'h et al.~(2005), who found $[(1+z_2)/(1+z_1)]^{3.9}$ between $z_1=0$ and $z_2=1$. The bolometric IR luminosity density at $z=1$ determined by Le Floc'h et al.~(2005; right-hand pointing triangle in Figure \ref{fig_lirdens}) is actually very close to the value we determine here. The difference appears to be mainly produced in the IR luminosity density at low redshifts: there has been a significant  evolution of the IR LF between redshifts $z=0$ and $z\sim0.2$.  
 
 Other symbols in Figure \ref{fig_lirdens} refer to different bolometric IR luminosity density estimations derived from different datasets: radio (Haarsma et al.~2000; downward-pointing triangle), sub-millimetre (Barger et al.~2000; diamond-like symbol) and the different fits made on mid-IR data by  P\'erez-Gonz\'alez et al.~(2005; asterisks). Our determinations of the IR luminosity densities are in good agreement with most of these previous works within the error bars. Our results exclude, however, the highest of the three estimations made by  P\'erez-Gonz\'alez et al.~(2005) at $z\gsim1$.

  Finally, in Figure \ref{fig_lirdens} we show the relative contributions of the LIRG and ULIRG populations to the total IR luminosity density, as a function of redshift. At $z\sim0$, $(28^{+11}_{-20})\%$ of the bolometric IR luminosity density is contained in LIRG and $<1\%$ in ULIRG. At $z=1$, we find that LIRG and ULIRG contribute $(61^{+4}_{-7})\%$ and $(16^{+11}_{-12})\%$, respectively, to the total IR luminosity density, in agreement with Le Floc'h et al.~(2005) within the error bars. By $z\sim2$, the contribution of LIRG and ULIRG become  $(47^{+13}_{-11})\%$ and  $(42^{+15}_{-22})\%$ of the total  budget, respectively.

  Using the Kennicutt formula $SFR=1.72 \times 10^{-10} \, L_{\rm IR}$, we can convert the bolometric  IR luminosity densities into star-formation rate densities at different redshifts. At $z=1$ and $z\sim2$, respectively, $\Omega_{\rm IR}=(1.2\pm0.2)\times 10^9$ and $(6.6^{+1.2}_{-1.0})\times 10^8 \, \rm L_\odot Mpc^{-3}$ translate into  star-formation rate densities $\delta_{\rm SFR} \approx (0.20\pm0.03)$ and $ (0.11\pm0.02)\, \rm  M_\odot \, yr^{-1} \, Mpc^{-3}$ (assuming a Salpeter initial mass function over stellar masses $M=(0.1-100)\, \rm M_\odot$). In Section \ref{sec_disc}, we make use of our current knowledge on stellar mass density evolution to discuss why these derived star-formation rate densities could not be much higher than this value at redshifts $1\lsim z \lsim 3$.

\section{Discussion}
\label{sec_disc}

  If  the IR LF for star-forming galaxies follows a unique law from the local Universe to high redshifts, then the results of our  LF determination will imply that there is a negative evolution in the overall number density of IR star-forming galaxies between $z\sim0$ and $z\sim2$.
We showed here the validity of a universal law to describe the IR LF at intermediate and bright luminosities, at different redshifts. Of course, one could argue that the faint-end of this LF is not sufficiently well constrained as to determine the number density of low-luminosity objects. Although a direct probe of the faint-end of the IR LF will require the capabilities of next-generation telescopes as the {\em James Webb Space Telescope (JWST)}, the stacking analysis of galaxies below the limits of our \tfm  survey appears to support our conclusion. The result of stacking analysis  suggests that  the faint-end slope of the IR LF at $z\sim2$ cannot be much higher than the value we considered here (and those usually considered in the literature at different redshifts).   
    
  In fact, an analogous situation is observed at other wavelengths. For example, Caputi et al.~(2006c) determined the evolution of the rest-frame $K_s$-band LF from $z=0$ to $z\sim2.5$. The depth of their survey ($K_s<21.5$ Vega mag) allowed them to properly constrain  this LF down to more than a magnitude below the turnover $M^\ast$ at $z=2$. These authors  found that a Schechter function with a same fixed slope is suitable to describe the $K_s$-band LF from the local Universe to high redshifts, within the limits of their survey. In this case, the ML analysis (which is in good agreement with the LF computed with the $1/V_{\rm max}$ method)   also indicates there is a negative density evolution of this LF with increasing redshift.

   The similarities between the evolutions of the $K_s$-band and \egm LF should not come as a surprise. The bright-end of the mid-IR LF at $z\sim2$ is mostly populated by massive $M\gsim 10^{11} \, \rm M_\odot$ galaxies (Caputi et al.~2006a). At redshift $z\sim1$, the mid-IR LF is dominated by LIRG, the majority of which are characterised by intermediate $\sim 10^{10}-10^{11} \, \rm M_\odot$ stellar masses (Hammer et al.~2005; Caputi et al.~2006a). Thus, the evolution in the number density of mid-IR galaxies above a given luminosity cut is related to the global evolution of galaxies above a given mass cut. 
  
  It should be clear that the aim of this discussion is to show how the results we find in this work are perfectly consistent with other observational evidence of galaxy evolution. This does not exclude, however, that the ultimate conclusion on the faint-ends of the $K_s$ and IR LF will only be achieved in the light of future extremely deep surveys.

 As we mentioned in Section \ref{sec_ldens}, the IR luminosity density associated with star-forming galaxies at  $z\sim2$  implies a star-formation rate density $(0.11\pm0.02) \, \rm  M_\odot \, yr^{-1} \, Mpc^{-3}$ (Kennicutt 1998).  Let us assume that this has been the average star-formation rate density between redshifts $z=1$ and $z=3$. In our assumed cosmology, the elapsed time between these redshifts is $\sim3.6 \, \rm Gyr$.  The stellar mass density formed during this period of time would be nearly $(4.0\pm0.7)\times 10^8 \, \rm M_\odot \, Mpc^{-3}$. If we consider that the  fraction of material recycled through stellar winds and supernovae could be roughly $50\%$, then the resulting mass density locked in stars would grow by  $\sim (2.0\pm0.4)\times 10^8 \, \rm M_\odot \, Mpc^{-3}$ between $z=3$ and $z=1$. This is actually  the growth of the stellar mass density that has been measured from different near-IR surveys at these redshifts (see Caputi et al.~2006c and references therein).  This result also shows that, unless the recycled fraction of material into the intestellar medium is much larger than $50\%$ between redshifts $z=1$ and $z=3$, then the average star-formation rate density along this period cannot very much exceed the value we find in this work  $\delta_{\rm SFR}\approx (0.11\pm0.02) \, \rm  M_\odot \, yr^{-1} \, Mpc^{-3}$ at $z\sim2$. Much higher star-formation rate densities only could be explained if a high fraction of the new formed stars were very massive, in which case they would not basically  contribute to the final stellar mass of the host galaxies.

 Considering  a star formation rate density $\delta_{\rm SFR}=(0.11\pm0.02) \, \rm  M_\odot \, yr^{-1} \, Mpc^{-3}$  strictly in the redshift range $1.7<z<2.3$ and assuming again a recycled fraction of 50\%, we derive that the total stellar mass density produced in this redshift interval is $(1.8\pm0.3)\times 10^7 \, \rm M_\odot \, Mpc^{-3}$. This is nearly 4\% of the total stellar mass density assembled at $z=0$ (i.e. $(4.9\pm0.1)\times 10^8 \, \rm M_\odot \, Mpc^{-3}$, as obtained by integrating the local stellar mass function of e.g. Cole et al.~2001). In the redshift interval $0.9<z<1.1$, our measured star formation rate density is $\delta_{\rm SFR}=(0.20\pm0.03) \, \rm  M_\odot \, yr^{-1} \, Mpc^{-3}$. With a recycled fraction of 50\%, this implies a growth in stellar mass density  of $(8.0\pm1.2)\times 10^7 \, \rm M_\odot \, Mpc^{-3}$. Thus, more than 15\% of the  present-day stellar mass density is being created in IR galaxies during the time elapsed between redshifts $z=0.9$ and $z=1.1$ (i.e. $\sim$0.8 Gyr).

  We found in this work that the number densities of ULIRG associated with star formation are very similar at redshifts $z=1$ and $\sim 2$. This suggests that the physical mechanism responsible for galaxies to enter a star-forming ULIRG phase is similarly efficient at these two redshifts. This result imposes strong constraints on IR-galaxy synthesis models. The origin of the ULIRG phase is usually associated with advanced gas-rich mergers  (Sanders \& Mirabel 1996). Thus, this phenomenon had to be comparably common for the production of powerful star-forming systems at redshifts $z=1$ and 2.

\section{Summary and Conclusions}
\label{sec_conc}

 In this work we have presented the IR LF of  \tfm-selected {\em Spitzer} galaxies at redshifts $z=1$ and $z\sim2$  in the GOODS fields. At $z\sim2$, we separately studied the LF for star-forming galaxies only and the total \egm LF for  star-forming galaxies and  AGN.  We then used a new calibration based on {\em Spitzer} star-forming galaxies to convert the rest-frame \egm into bolometric IR luminosities of the star-forming galaxies in our sample. This allowed us to compute the bolometric IR LF and obtain an estimate of the IR luminosity densities at $z=1$ and $z\sim2$.

 We found that the rest-frame \egm LF for star-forming galaxies at  $z=1$  and $z\sim2$ is well described  by a double-exponential law which has evolved from $z\sim0$. Between $z\sim0$ and $z=1$, there is a strong luminosity evolution and the number density of $\log_{10}(\nu L_{\nu}^{8 \,\rm \mu m})>10.5$ increases by a factor $>20$. The characteristic luminosity $L^\ast$ of the rest-frame \egm LF continues increasing up to redshift $z\sim2$, but, at this redshift, the number density of $\log_{10}(\nu L_{\nu}^{8 \,\rm \mu m})>10.5$ galaxies is smaller than the density at $z=1$.
 This certainly does not mean that the contribution of IR galaxies has been less important at high redshifts. The rest-frame \egm luminosity density  at $z\sim2$ is still $\sim2.3$ times larger than the corresponding luminosity density at $z\sim0$, but only half the value at $z=1$.

At $z\sim2$, the inclusion of AGN mainly affects the bright end of the IR LF.  The bright end of the total rest-frame \egm LF for star-forming galaxies and AGN is correctly reproduced by a power-law which accounts for the excess of bright sources. AGN only produce  $\sim 17\%$  of  the  total rest-frame \egm luminosity density at $z\sim2$.

The quasi-linear relation between rest-frame \egm and bolometric IR luminosities for star-forming galaxies makes that the bolometric IR LF  is well-described by a similar law as the   rest-frame \egm LF at the same redshift. The  characteristic luminosity $L^\ast_{\rm IR}$ of the bolometric IR LF for star-forming galaxies at $z\sim2$ is close to $\sim 10^{12} \, \rm L_\odot$, i.e.  the limiting luminosity between the LIRGs and ULIRGs. As the luminosity density is mainly governed by the turnover of the LF, the value of $L^\ast_{\rm IR}$ results in roughly similar contributions of LIRGs and ULIRGs to the IR luminosity density. These two populations altogether account for $\sim 90\%$ of the total IR luminosity density associated with star formation at $z\sim2$.

Finally, we discussed the possibility that the total IR luminosity and corresponding star-formation rate density  estimated in this work could have been significantly different at any redshift between $z=1$ and $z=3$. Constraints from near-IR surveys suggest that the stellar mass density built up by galaxies at this epoch would be in contradiction with average star-formation rate densities much larger than our estimated value (unless a much higher proportion of very massive stars were created in the past). Our results appear, then, to be consistent with this other observational evidence of galaxy evolution.

%
\acknowledgments

This paper is based on observations made with the {\em Spitzer} Observatory, which is operated by the Jet Propulsion Laboratory, California Institute of Technology, under NASA contract 1407. Also based on observations made with the Advanced Camera for Surveys on board the Hubble Space Telescope operated by NASA/ESA and with the Infrared Spectrometer and Array Camera on the `Antu' Very Large Telescope operated by the European Southern Observatory in Cerro Paranal, Chile,  and  form part of the publicly available GOODS datasets.  We thank the GOODS teams for providing reduced data products.

 We are grateful to H\'ector Flores and Fran{\c c}ois Hammer, for providing us additional spectroscopic redshifts for the GOODS/CDFS; Jiasheng Huang, for sending us the results of his  \egm LF at $z\sim0$ before publication;  Dieter Lutz and Pablo P\'erez-Gonz\'alez, for providing us some of their published results in  electronic format. We thank the anonymous referee for helpful comments and suggestions. KIC and GL thank the Infrared Processing and Analysis Center (IPAC) at Caltech for hospitality while part of this work has been done. KIC aknowledges CNES and CNRS funding.

%

\clearpage

%
\begin{deluxetable}{rc} 
\tablewidth{0mm} 
\tablecaption{The rest-frame \egm LF for star-forming galaxies at $z=1$ obtained with the $1/V_{\rm max}$ method. 
\label{tab_egm_01}} 
\tablehead{  
\colhead{$\log_{10}(\nu L_{\nu}^{8 \,\rm \mu m})$} &
\colhead{$\log_{10}\Phi \,(\rm Mpc^{-3} \, dex^{-1})$} 
} 
\startdata 
10.60 & $-2.55^{+0.06}_{-0.08}$ \\
10.80 & $-2.66^{+0.07}_{-0.07}$ \\
11.00 & $-2.97^{+0.09}_{-0.10}$ \\
11.20 & $-3.30^{+0.12}_{-0.15}$ \\
11.40 & $-4.12^{+0.25}_{-0.54}$ \\ 
\enddata 
\end{deluxetable}

%
\begin{deluxetable}{lccrr} 
\tablewidth{0mm} 
\tablecaption{Parameter values characterising the rest-frame \egm LF for star-forming galaxies at $z=1$.
\label{tab_egm_02}} 
\tablehead{  
\colhead{Functional form} &
\colhead{$\alpha$} &
\colhead{$\sigma$}&
\colhead{$\nu L_{\nu}^{\ast \, 8 \,\rm \mu m} \, (\rm L_\odot)$}&
\colhead{$\Phi^\ast \, \rm (Mpc^{-3} dex^{-1})$}
} 
\startdata 
Double-exp (eq. \ref{eq-dexp}) & 1.2 (fixed)& 0.36 (fixed)& $(3.55^{+0.52}_{-0.40})\times 10^{10}$ & $(3.95^{+0.50}_{-0.49})\times 10^{-3}$ \\
Double-exp (eq. \ref{eq-dexp}) & 1.2 (fixed)& $0.20^{+0.11}_{-0.07}$ (free)& $(1.10^{+0.99}_{-0.64})\times 10^{11}$ & $(2.54^{+0.60}_{-0.35})\times 10^{-3}$ \\
Schechter (eq. \ref{eq-sch}) & 1.2 (fixed)& --- & $(7.2^{+0.9}_{-0.7})\times 10^{10}$ & $(3.88^{+0.46}_{-0.41})\times 10^{-3}$\\ 
\enddata 
\end{deluxetable}

%
\begin{deluxetable}{rc} 
\tablewidth{0mm} 
\tablecaption{The rest-frame \egm LF for star-forming galaxies at $z\sim2$ obtained with the $1/V_{\rm max}$ method. 
\label{tab_egm_03}} 
\tablehead{  
\colhead{$\log_{10}(\nu L_{\nu}^{8 \,\rm \mu m})$} &
\colhead{$\log_{10}\Phi \,(\rm Mpc^{-3} \, dex^{-1})$} 
} 
\startdata 
11.09 & $-3.34^{+0.06}_{-0.09}$ \\
11.29 & $-3.49^{+0.09}_{-0.08}$ \\
11.49 & $-3.88^{+0.18}_{-0.13}$ \\
11.69 & $-4.58^{+0.29}_{-0.38}$ \\
 
\enddata 
\end{deluxetable}

%
\begin{deluxetable}{lccrr} 
\tablewidth{0mm} 
\tablecaption{Parameter values characterising the rest-frame \egm LF for star-forming galaxies at $z\sim2$.
\label{tab_egm_04}} 
\tablehead{  
\colhead{Functional form} &
\colhead{$\alpha$} &
\colhead{$\sigma$}&
\colhead{$\nu L_{\nu}^{\ast \, 8 \,\rm \mu m} \, (\rm L_\odot)$}&
\colhead{$\Phi^\ast \, \rm (Mpc^{-3} dex^{-1})$}
} 
\startdata 
Double-exp (eq. \ref{eq-dexp}) & 1.2 (fixed) & 0.36 (fixed) & $(8.3^{+1.5}_{-1.1})\times 10^{10}$ & $(9.0^{+2.1}_{-1.7})\times 10^{-4}$ \\
 Schechter (eq. \ref{eq-sch}) & 1.2 (fixed) & --- & $(1.62^{+0.20}_{-0.21})\times 10^{11}$ & $(9.3^{+2.1}_{-1.3})\times 10^{-4}$\\ 
\enddata 
\end{deluxetable}

%
\begin{deluxetable}{clrrr} 
\tablewidth{0mm} 
\tablecaption{Number densities of galaxies with rest-frame  \nLnegm above different luminosity cuts at different redshifts. These number densities have been obtained by integrating the functional form appearing in the second column and are expressed in units of $\rm Mpc^{-3}$. DE stands for double-exponential.
\label{tab_numd}} 
\tablehead{ 
\colhead{Redshift} & 
\colhead{Functional form} &
\colhead{$\log_{10}(\nu L_\nu^{\rm 8 \, \mu m})>10.5$} &
\colhead{$>11.0$}&
\colhead{$>11.5$}
} 
\startdata 
$z\sim0$ & DE  ($\sigma=0.36$) &  $(4.8\pm 0.4)\times10^{-5}$ & $(6.7\pm0.9)\times10^{-7}$ & $(1.4\pm0.3)\times10^{-9}$\\
 & & & & \\
$z=1$ & DE ($\sigma=0.36$) &  $(1.1\pm0.1)\times10^{-3}$ & $(1.8\pm0.3)\times10^{-4}$ & $(6.7\pm2.0)\times10^{-6}$\\  
  & DE  ($\sigma=0.20$) &  $(1.1\pm0.1)\times10^{-3}$ & $(1.7\pm0.3)\times10^{-4}$ & $(2.1\pm1.0)\times10^{-6}$\\
  & Schechter   &  $(1.1\pm0.1)\times10^{-3}$ & $(1.7\pm0.3)\times10^{-4}$ & $(2.9\pm1.5)\times10^{-6}$\\
  & & & & \\
$z\sim2$ & DE  ($\sigma=0.36$) &  $(5.7\pm0.5)\times10^{-4}$ & $(1.7\pm0.2)\times10^{-4}$ & $(2.0\pm0.4)\times10^{-5}$\\  
  & Schechter   &  $(5.8\pm0.4)\times10^{-4}$ & $(1.7\pm0.2)\times10^{-4}$ & $(1.7\pm0.4)\times10^{-5}$\\
\enddata 
\end{deluxetable}

%
\begin{deluxetable}{rc} 
\tablewidth{0mm} 
\tablecaption{The rest-frame \egm LF for all galaxies at $z\sim2$ obtained with the $1/V_{\rm max}$ method. 
\label{tab_egm_05}} 
\tablehead{  
\colhead{$\log_{10}(\nu L_{\nu}^{8 \,\rm \mu m})$} &
\colhead{$\log_{10}\Phi \,(\rm Mpc^{-3} \, dex^{-1})$} 
} 
\startdata 
11.09 & $-3.26^{+0.05}_{-0.09}$ \\
11.29 & $-3.41^{+0.08}_{-0.07}$ \\
11.49 & $-3.83^{+0.18}_{-0.12}$ \\
11.69 & $-4.28^{+0.15}_{-0.29}$ \\
11.99 & $-4.88^{+0.29}_{-0.34}$ \\
\enddata 
\end{deluxetable}

%
\begin{deluxetable}{crccrr} 
\tablewidth{0mm} 
\tablecaption{Parameter values characterising the bolometric IR LF for star-forming galaxies at $z=1$ and $z\sim2$.
\label{tab_bol}} 
\tablehead{ 
\colhead{Redshift} &
\colhead{Functional form} &
\colhead{$\alpha$} &
\colhead{$\sigma$}&
\colhead{$L_{\rm IR}^{\ast} \, (\rm L_\odot)$}&
\colhead{$\Phi^\ast \, \rm (Mpc^{-3} dex^{-1})$}
} 
\startdata
$z=1$  & Double-exp (eq. \ref{eq-dexp}) & 1.2 & 0.39 (fixed) & $(2.5^{+0.4}_{-0.3})\times 10^{11}$ & $(4.0^{+0.6}_{-0.5})\times 10^{-3}$\\
$z\sim2$  & Double-exp (eq. \ref{eq-dexp}) & 1.2 & 0.39 (fixed) & $(6.3^{+1.1}_{-0.9})\times 10^{11}$ & $(9.2^{+2.2}_{-1.7})\times 10^{-4}$\\
\enddata 
\end{deluxetable}

%
\begin{deluxetable}{clrrr} 
\tablewidth{0mm} 
\tablecaption{Number densities of star-forming LIRG and ULIRG at different redshifts. These number densities have been obtained by integrating the functional form appearing in the second column and are expressed in units of $\rm Mpc^{-3}$. DE stands for double-exponential.
\label{tab_bolnd}} 
\tablehead{ 
\colhead{Redshift} & 
\colhead{Functional form} &
\colhead{$\log_{10}(L_{\rm bol.}^{\rm IR})>11$} &
\colhead{LIRG}&
\colhead{ULIRG}
} 
\startdata 
$z\sim0$ & DE ($\sigma=0.39$) &  $(4.1\pm0.3)\times10^{-4}$ & $(4.1\pm0.3)\times10^{-4}$ & $(3.9\pm0.7)\times10^{-7}$\\
$z=1$ & DE  ($\sigma=0.39$) &  $(2.6\pm0.1)\times10^{-3}$ & $(2.5\pm0.2)\times10^{-3}$ & $(1.2\pm0.2)\times10^{-4}$\\  
 
$z\sim2$ & DE  ($\sigma=0.39$) &  $(1.1\pm0.1)\times10^{-3}$ & $(9.5\pm1.5)\times10^{-4}$ & $(1.5\pm0.2)\times10^{-4}$\\
\enddata 
\end{deluxetable}

\clearpage

%

%

%

%

%

%

\clearpage

%

%

%

%

\clearpage

%

%

%

%

%

\end{document}